\documentclass[preprintnumbers]{revtex4}
\pdfoutput=1
\usepackage{eurosym}
\usepackage{amsfonts}
\usepackage{amsmath}
\usepackage{hyperref}
\usepackage{amssymb}
\usepackage[english]{babel}
\usepackage{graphicx}
\usepackage{epsfig}
\usepackage{bm}
\usepackage{color}
\usepackage{longtable}
\usepackage{verbatim}
\usepackage{longtable}
\usepackage{multirow}
\usepackage[utf8]{inputenc}

\newcommand{\mathsym}[1]{{}}
\input{tcilatex}
\begin{document}

\title{$\Delta \left( 27\right) $ flavor singlet-triplet Higgs
model for fermion masses and mixings.}

\author{A. E. C\'{a}rcamo Hern\'{a}ndez${}^{a}$}
\email{antonio.carcamo@usm.cl}
\author{Juan Carlos G\'{o}mez-Izquierdo${}^{b,c,d}$}
\email{jcgizquierdo1979@gmail.com}
\author{Sergey Kovalenko${}^{a}$}
\email{sergey.kovalenko@usm.cl}
\author{Myriam Mondrag\'on${}^{c}$}
\email{myriam@fisica.unam.mx}
\affiliation{$^{{a}}$Universidad T\'{e}cnica Federico Santa Mar\'{\i}a\\
and Centro Cient\'{\i}fico-Tecnol\'{o}gico de Valpara\'{\i}so\\
Casilla 110-V, Valpara\'{\i}so, Chile,\\
$^{{b}}$Departamento de F\'isica, Centro de Investigaci\'on y de Estudios
Avanzados del I. P. N.,\\
Apdo. Post. 14-740, 07000, Ciudad de M\'exico, M\'exico,\\
$^{{c}}$Departamento de F\'{\i}sica Te\'{o}rica, Instituto de F\'{\i}sica,\\
Universidad Nacional Aut\'onoma de M\'{e}xico\\
A.P. 20-364 01000, M\'{e}xico D.F,\\
$^{{d}}$Centro de Estudios Cient\'ificos y Tecnol\'ogicos No 16, \\
Instituto Polit\'ecnico Nacional, Pachuca: Ciudad del Conocimiento y la Cultura,\\ 
Carretera Pachuca Actopan km 1+500, San Agust\'in Tlaxiaca, Hidalgo, M\'exico.
}
\date{\today }

\begin{abstract}
We propose a multiscalar singlet extension of the
singlet-triplet Higgs model capable of explaining 
 the SM fermion mass spectrum and mixing parameters.
 Our model is based on the $\Delta \left( 27\right) $ family symmetry, supplemented by cyclic symmetries, which are spontaneously broken thus yielding 
the
observed hierarchy of the SM charged fermion masses and quark mixing angles.
The masses of the light active neutrinos are produced by type-II seesaw mechanism mediated by the neutral component
of the $SU(2)_{L}$ scalar triplet. The model symmetries lead to the extended 
Gatto-Sartori-Tonin relations between the quark masses and mixing angles.
\end{abstract}

\maketitle

\section{Introduction}

Despite its great experimental success, the Standard Model (SM) does not answer several fundamental questions such as, for instance, the number of fermion
families and the observed pattern of fermion masses and mixing angles. While the quark mixing angles are very small, thus implying a quark mixing matrix close to the identity matrix, two of the leptonic mixing angles are large, and one is small, i.e, of the order of the Cabbibo angle, indicating that the leptonic mixing matrix is significantly different from the identity matrix. This aforementioned experimental fact suggests
a different kind of New Physics 
in the neutrino sector in comparison with the quark sector.
The Daya Bay \cite{An:2012eh}, T2K 
\cite{Abe:2011sj}, MINOS \cite{Adamson:2011qu}, Double CHOOZ \cite%
{Abe:2011fz} and RENO \cite{Ahn:2012nd} neutrino oscillation experiments,
have brought clear evidence that at least two of the light active neutrinos
have non-vanishing masses. These experiments have provided important
constraints on the neutrino mass squared splittings and leptonic mixing
parameters \cite{deSalas:2017kay}. On the other hand,
the SM does is unable to explain the large hierarchy of fermion masses, which is extended over a range of 
about thirteen  orders of magnitude, from the neutrino mass scale up to the top quark mass.
%
%
This, the so-called ``flavor puzzle'' which is left without an explanation by the SM,  
motivates the introduction of
various
SM extensions with larger scalar and/or fermion
sectors with extended gauge group and discrete flavour
symmetries in order to get viable and predictive fermion mass matrix textures that explain the SM fermion mass spectrum and fermionic mixing parameters.
 Using discrete
family symmetries in several theories corresponding to extensions of the Standard Model allow to successfully explain the observed pattern of charged fermion masses and mixing angles (recent
reviews on discrete flavor groups are provided in Refs. \cite%
{Ishimori:2010au,Altarelli:2010gt,King:2013eh,King:2014nza,King:2017guk,Petcov:2017ggy}). In this line of thought, several
 discrete groups have been employed in those extensions of the SM, such as $S_{3}$ \cite{Gerard:1982mm,Kubo:2003iw,Kubo:2003pd,Kobayashi:2003fh,Chen:2004rr,Mondragon:2007af,Mondragon:2008gm,Bhattacharyya:2010hp, Dong:2011vb,Dias:2012bh,Meloni:2012ci,Canales:2012dr,Canales:2013cga,Ma:2013zca,Kajiyama:2013sza,Hernandez:2013hea, Ma:2014qra,Hernandez:2014vta,Hernandez:2014lpa,Gupta:2014nba,Hernandez:2015dga,Hernandez:2015zeh,Hernandez:2015hrt,Hernandez:2016rbi,CarcamoHernandez:2016pdu,Arbelaez:2016mhg,Gomez-Izquierdo:2017rxi,Cruz:2017add,Ma:2017trv,Espinoza:2018itz,Garces:2018nar,CarcamoHernandez:2018vdj,Gomez-Izquierdo:2018jrx}%
, $A_{4}$ \cite%
{Ma:2001dn,He:2006dk,Chen:2009um,Ahn:2012tv,Memenga:2013vc,Felipe:2013vwa,Varzielas:2012ai, Ishimori:2012fg,King:2013hj,Hernandez:2013dta,Babu:2002dz,Altarelli:2005yx,Gupta:2011ct,Morisi:2013eca, Altarelli:2005yp,Kadosh:2010rm,Kadosh:2013nra,delAguila:2010vg,Campos:2014lla,Vien:2014pta,Joshipura:2015dsa,Hernandez:2015tna,Chattopadhyay:2017zvs,CarcamoHernandez:2017kra,Ma:2017moj,CentellesChulia:2017koy,Bjorkeroth:2017tsz,Srivastava:2017sno,Belyaev:2018vkl,CarcamoHernandez:2018aon,Srivastava:2018ser,delaVega:2018cnx,Pramanick:2019qpg}%
, $S_{4}$ \cite%
{Patel:2010hr,Morisi:2011pm,Mohapatra:2012tb,BhupalDev:2012nm,Varzielas:2012pa,Ding:2013hpa,Ishimori:2010fs,Ding:2013eca,Hagedorn:2011un,Campos:2014zaa,Dong:2010zu,VanVien:2015xha,deAnda:2017yeb,deAnda:2018oik,CarcamoHernandez:2019eme}, $D_{4}$ \cite%
{Frampton:1994rk,Grimus:2003kq,Grimus:2004rj,Frigerio:2004jg,Adulpravitchai:2008yp,Ishimori:2008gp,Hagedorn:2010mq,Meloni:2011cc,Vien:2013zra}%
, $Q_{6}$ \cite%
{Babu:2004tn,Kajiyama:2005rk,Kajiyama:2007pr,Kifune:2007fj,Babu:2009nn,
Kawashima:2009jv,Kaburaki:2010xc,Babu:2011mv,Araki:2011zg, Gomez-Izquierdo:2013uaa,Gomez-Izquierdo:2017med}%
, $T_{7}$ \cite%
{Luhn:2007sy,Hagedorn:2008bc,Cao:2010mp,Luhn:2012bc,Kajiyama:2013lja,Bonilla:2014xla,Vien:2014gza, Vien:2015koa,Hernandez:2015cra,Arbelaez:2015toa}%
, $T_{13}$ \cite{Ding:2011qt,Hartmann:2011dn,Hartmann:2011pq,Kajiyama:2010sb}%
, $T^{\prime }$ \cite%
{Aranda:2000tm,Feruglio:2007uu,Sen:2007vx,Aranda:2007dp,Chen:2007afa,Eby:2008uc,Frampton:2008bz,Frampton:2008ep,Eby:2009ii,Frampton:2009fw,Eby:2011ph,Eby:2011qa,Chen:2011tj,Frampton:2013lva,Chen:2013wba,Girardi:2013sza}%
, $\Delta (27)$ \cite%
{Branco:1983tn,deMedeirosVarzielas:2006fc,Ma:2007wu,Varzielas:2012nn,Bhattacharyya:2012pi,Ferreira:2012ri,Ma:2013xqa,Nishi:2013jqa,Varzielas:2013sla,Aranda:2013gga,Harrison:2014jqa,Ma:2014eka,Abbas:2014ewa,Abbas:2015zna,Varzielas:2015aua,Bjorkeroth:2015uou,Chen:2015jta,Vien:2016tmh,Hernandez:2016eod,CarcamoHernandez:2017owh,deMedeirosVarzielas:2017sdv,Bernal:2017xat,CarcamoHernandez:2018iel,deMedeirosVarzielas:2018vab,CarcamoHernandez:2018hst}%
, $\Delta(54)$ \cite{Carballo-Perez:2016ooy}, $\Delta(96)$ \cite{King:2012in,King:2013vna,Ding:2014ssa}, $\Delta(6N^2)$ \cite{Ishimori:2014jwa,King:2014rwa,Ishimori:2014nxa}, $\Sigma(26\times 3)$ \cite{Rong:2016cpk}, $\Sigma(72\times 3)$ \cite{Krishnan:2018tja} and $A_{5}$ \cite%
{Everett:2008et,Feruglio:2011qq,Cooper:2012bd,Varzielas:2013hga,Gehrlein:2014wda,Gehrlein:2015dxa,DiIura:2015kfa,Ballett:2015wia,Gehrlein:2015dza,Turner:2015uta,Li:2015jxa}.

An extra motivation to use discrete flavor symmetries is that these can arise from the underlying theory, e. g.,
string theory or compactification via orbifolding. In particular, from
the heterotic orbifold models 
one can generate the $D_{4}$ and $\Delta(54)$ flavor symmetries \cite%
{Kobayashi:2006wq,Kobayashi:2008ih,Abe:2009vi,BerasaluceGonzalez:2011wy,Beye:2014nxa}%
. Furthermore, magnetized/intersecting D-brane models can generate the $%
\Delta(27)$ flavor symmetry \cite%
{Kobayashi:2006wq,Kobayashi:2008ih,Abe:2009vi,BerasaluceGonzalez:2011wy,Beye:2014nxa}. 
%

In this paper we propose a predictive multiscalar singlet extension of the singlet-triplet Higgs
 model, capable of explaining the current SM fermion mass spectrum and fermionic mixing parameters. In our model we use 
$\Delta\left(27\right)$ family symmetry supplemented by other auxiliary cyclic symmetries, whose spontaneous breaking at very large energy scale, produces predictive and viable textures for the fermion sector, which in a natural benchmark scenario with just two free quark sector parameters, allows to reasonably reproduce the
experimental values of the ten physical observables of the quark sector. 
The model symmetries yield extended Gatto-Sartori-Tonin relations between the quark masses and mixing angles consistent with the low energy quark flavor data. In our model the masses of the light active neutrinos are generated from a type-II seesaw mechanism mediated
by the neutral component of the $SU(2)_L$ scalar triplet. The experimental values for the physical observables of the lepton sector are also successfully reproduced for both normal and inverted neutrino mass hierarchy. We use the $\Delta(27)$ discrete group, since it is the smallest non trivial group of the type $\Delta(3n^2)$ and is isomorphic to the semi-direct product group $(Z_{3}^{\prime }\times Z_{3}^{\prime \prime })\rtimes Z_{3}$. 

The layout of the remainder of the paper is as follows. In section \ref{model} we
describe the model-setup. In section \ref{quarksector} we present the
implications of our model for the quark sector observables. Section \ref%
{leptonsector} discusses the masses and mixings in the lepton sector. Our conclusions are provided in section %
\ref{conclusions}. A concise description of the $\Delta\left(27\right)$ discrete group is presented in Appendix \ref{A}.

\section{The model}

\label{model} We propose an extension of the singlet-triplet Higgs model
where the full symmetry $\mathcal{G}$ features the following spontaneous symmetry breaking chain:
\begin{eqnarray}
&&\mathcal{G}=SU(3)_{C}\times SU\left( 2\right) _{L}\times U\left( 1\right)
_{Y}\times \Delta \left( 27\right) \times Z_{16}\times Z_{24}  \notag \\
&&\hspace{35mm}\Downarrow \Lambda _{int}  \notag \\[0.12in]
&&\hspace{15mm}SU(3)_{C}\times SU\left( 2\right) _{L}\times U\left( 1\right)
_{Y}  \notag \\[0.12in]
&&\hspace{35mm}\Downarrow v  \notag \\[0.12in]
&&\hspace{15mm}SU(3)_{C}\times U\left( 1\right) _{Q}
\end{eqnarray}%
where the scale of spontaneous breaking of the $\Delta \left( 27\right) \times Z_{16}\times Z_{24}$ discrete group, namely $\Lambda
_{int}$ is assumed to be much larger than the Fermi scale, thus implying the hierarchy $\Lambda
_{int}\gg v$, where $v=246$ GeV. 

The fermion assignments under the group $\Delta \left( 27\right) \times
Z_{16}\times Z_{24}$ are: 
\begin{eqnarray}
q_{1L} &\sim &\left( \mathbf{1}_{\mathbf{0,2}}\mathbf{,}e^{-\frac{3i\pi }{8}%
},1\right) ,\hspace{1.5cm}q_{2L}\sim \left( \mathbf{1}_{\mathbf{0,2}}\mathbf{%
,}e^{-\frac{i\pi }{4}},1\right) ,\hspace{1.5cm}q_{3L}\sim \left( \mathbf{1}_{%
\mathbf{0,0}},1,1\right) ,  \notag \\
u_{1R} &\sim &\left( \mathbf{1}_{\mathbf{0,0}},e^{\frac{5i\pi }{8}},1\right)
,\hspace{1.5cm}u_{2R}\sim \left( \mathbf{1}_{\mathbf{1,1}}\mathbf{,}e^{\frac{%
i\pi }{4}},1\right) ,\hspace{1.5cm}u_{3R}\sim \left( \mathbf{1}_{\mathbf{0,0}%
}\mathbf{,}1,1\right) ,  \notag \\
d_{1R} &\sim &\left( \mathbf{1}_{\mathbf{0,1}},e^{\frac{5i\pi }{8}},1\right)
,\hspace{1.5cm}d_{2R}\sim \left( \mathbf{1}_{\mathbf{2,0}},e^{\frac{3i\pi }{8%
}},e^{\frac{i\pi }{3}}\right) ,\hspace{1.5cm}d_{3R}\sim \left( \mathbf{1}_{%
\mathbf{0,0}},e^{\frac{3i\pi }{8}},-1\right) ,  \notag \\
l_{L} &\sim &\left( \mathbf{3},1,1\right) ,\hspace{1.5cm}l_{1R}\sim \left( 
\mathbf{1}_{\mathbf{0,0}},e^{\frac{7i\pi }{8}},ie^{-\frac{7i\pi }{3}}\right)
,\hspace{1.5cm}l_{2R}\sim \left( \mathbf{1}_{\mathbf{1,0}},e^{\frac{4i\pi }{8%
}},e^{-\frac{4i\pi }{3}}\right)  \notag \\
l_{3R} &\sim &\left( \mathbf{1}_{\mathbf{2,0}},e^{\frac{i\pi }{8}},-ie^{-%
\frac{i\pi }{3}}\right) .
\end{eqnarray}

The scalar sector of the model is composed of an SM Higgs doublet $\phi $,
an $SU\left( 2\right) $ scalar triplet $\Delta $ with lepton number equal to 
$-2$ and hypercharge equal to $1$ and eighteen scalar SM singlets, i.e, $S$,$\ \sigma $,$\ \rho $,$\
\eta $,$\ \tau $,$\ \chi _{1}$, $\chi _{2}$,$\ \varphi _{1}$, $\varphi _{2}$%
, $\xi _{i}$, $\zeta _{i}$, $\Phi_{i}$ ($i=1,2,3$). 
Of these fifteen
only $S$ has a non-vanishing lepton number, which is set to 
$2$, to allow the quartic scalar interaction $S\phi ^{\dagger
}\Delta \phi $.  Then the $S$ is required to have vanishing hypercharge.
%
%
The $\Delta \left( 27\right) \times Z_{16}\times Z_{24}$
assignments for the scalar sector are: 
\begin{eqnarray}
\phi &\sim &\left( \mathbf{1}_{\mathbf{0,0}}\mathbf{,}1\mathbf{,}1\right) ,%
\hspace{1.5cm}S\sim \left( \mathbf{1}_{\mathbf{0,0}}\mathbf{,}1\mathbf{,}%
1\right) ,\hspace{1.5cm}\sigma \sim \left( \mathbf{1}_{\mathbf{0,1}},e^{-%
\frac{i\pi }{8}},1\right) ,\hspace{1.5cm}\rho \sim \left( \mathbf{1}_{%
\mathbf{2,1}}\mathbf{,}e^{-\frac{i\pi }{8}},1\right) ,  \notag \\
\eta &\sim &\left( \mathbf{1}_{\mathbf{0,0}},e^{-\frac{i\pi }{8}},e^{\frac{%
i\pi }{3}}\right) ,\hspace{1.5cm}\tau \sim \left( \mathbf{1}_{\mathbf{1,0}}%
\mathbf{,}e^{-\frac{i\pi }{8}},i^{\frac{1}{2}}\right) ,\hspace{1.5cm}\chi
_{1}\sim \left( \mathbf{1}_{\mathbf{1,0}}\mathbf{,}e^{-\frac{i\pi }{8}},e^{-%
\frac{i\pi }{12}}\right)  \notag \\
\chi _{2} &\sim &\left( \mathbf{1}_{\mathbf{1,1}}\mathbf{,}e^{-\frac{i\pi }{8%
}},e^{-\frac{i\pi }{12}}\right) ,\hspace{1.5cm}\xi \sim \left( \mathbf{3}%
,1,1\right) ,\hspace{1.5cm}\zeta \sim \left( \mathbf{3},-1,e^{-\frac{2i\pi }{3}}\right) ,%
\hspace{1.5cm}\Delta \sim \left( \mathbf{1}_{\mathbf{0,0}}\mathbf{,}1%
\mathbf{,}1\right) ,  \notag \\
\varphi _{1} &\sim &\left( \mathbf{1}_{\mathbf{0,0}},1,-i\right) ,\hspace{%
1.5cm}\varphi _{2}\sim \left( \mathbf{1}_{\mathbf{2,0}},-1,i^{\frac{1}{2}%
}\right)
\end{eqnarray}

\begin{table}[tbp]
\begin{tabular}{|c|c|c|c|c|c|c|c|c|c|c|c|c|c|}
\hline\hline
& $q_{1L}$ & $q_{2L}$ & $q_{3L}$ & $u_{1R}$ & $u_{2R}$ & $u_{3R}$ & $d_{1R}$
& $d_{2R}$ & $d_{3R}$ & $l_{L}$ & $l_{1R}$ & $l_{2R}$ & $l_{3R}$ \\ \hline
$\Delta\left(27\right)$ & $\mathbf{1}_{\mathbf{0,2}}$ & $\mathbf{1}_{\mathbf{%
0,2}}$ & $\mathbf{1}_{\mathbf{0,0}}$ & $\mathbf{1}_{\mathbf{0,0}}$ & $%
\mathbf{1}_{\mathbf{2,2}}$ & $\mathbf{1}_{\mathbf{0,0}}$ & $\mathbf{1}_{%
\mathbf{0,1}}$ & $\mathbf{1}_{\mathbf{2,0}}$ & $\mathbf{1}_{\mathbf{0,0}}$ & 
$\mathbf{3}$ & $\mathbf{1}_{\mathbf{0,0}}$ & $\mathbf{1}_{\mathbf{1,0}}$ & $%
\mathbf{1}_{\mathbf{2,0}}$ \\ \hline
$Z_{16}$ & $e^{-\frac{3i\pi }{8}}$ & $e^{-\frac{i\pi }{4}}$ & $1$ & $e^{%
\frac{5i\pi }{8}}$ & $e^{-\frac{i\pi }{4}}$ & $1$ & $e^{\frac{5i\pi }{8}}$ & $%
e^{\frac{3i\pi }{8}}$ & $e^{\frac{3i\pi }{8}}$ & $1$ & $e^{\frac{7i\pi }{8}}$
& $e^{\frac{4i\pi }{8}}$ & $e^{\frac{i\pi }{8}}$ \\ \hline
$Z_{24}$ & $1$ & $1$ & $1$ & $1$ & $-1$ & $1$ & $1$ & $e^{\frac{i\pi }{3}}$ & 
$-1$ & $1$ & $ie^{-\frac{7i\pi }{3}}$ & $e^{-\frac{4i\pi }{3}}$ & $-ie^{-%
\frac{i\pi }{3}}$ \\ \hline\hline
\end{tabular}%
\caption{Fermion assignments under $\Delta\left(27\right)\times Z_{16}\times
Z_{24}$.}
\label{ta:fermions}
\end{table}

\begin{table}[tbp]
\begin{tabular}{|c|c|c|c|c|c|c|c|c|c|c|c|c|c|}
\hline\hline
& $\phi$ & $\Delta$ & $\sigma$ & $\rho$ & $\eta$ & $\tau$ & $\chi_1$ & $%
\chi_2$ & $\xi$ & $\zeta$ & $\Phi$ & $\varphi_1$ & $\varphi_2$ \\ \hline
$\Delta\left(27\right)$ & $\mathbf{1}_{\mathbf{0,0}}$ & $\mathbf{1}_{\mathbf{%
0,0}}$ & $\mathbf{1}_{\mathbf{0,1}}$ & $\mathbf{1}_{\mathbf{2,1}}$ & $%
\mathbf{1}_{\mathbf{0,0}}$ & $\mathbf{1}_{\mathbf{1,0}}$ & $\mathbf{1}_{%
\mathbf{1,0}}$ & $\mathbf{1}_{\mathbf{1,1}}$ & $\mathbf{3}$ & $\mathbf{3}$ & $\mathbf{3}$ & 
$\mathbf{1}_{\mathbf{0,0}}$ & $\mathbf{1}_{\mathbf{2,0}}$ \\ \hline
$Z_{16}$ & $1$ & $1$ & $e^{-\frac{i\pi }{8}}$ & $e^{-\frac{i\pi }{8}}$ & $%
e^{-\frac{i\pi }{8}}$ & $e^{-\frac{i\pi }{8}}$ & $e^{-\frac{i\pi }{8}}$ & $%
e^{-\frac{i\pi }{8}}$ & $1$ & $-1$ & $-1$ & $1$ & $-1$ \\ \hline
$Z_{24}$ & $1$ & $1$ & $1$ & $1$ & $e^{\frac{i\pi }{3}}$ & $i^{\frac{1}{2}}$
& $e^{-\frac{i\pi }{12}}$ & $e^{-\frac{i\pi }{12}}$ & $1$ & $e^{-\frac{2i\pi }{3}}$ & $e^{-\frac{2i\pi }{3}}$ & $-i$ & $%
i^{\frac{1}{2}}$ \\ \hline
\end{tabular}%
\caption{Scalar assignments under $\Delta\left(27\right)\times Z_{16}\times
Z_{24}$. }
\label{ta:fermions}
\end{table}

We decompose the scalar fields 
around their VEVs as 
\begin{equation}
\label{eq:VEVs-1}
\phi =\left( 
\begin{array}{c}
\pi ^{+} \\ 
\frac{1}{\sqrt{2}}\left( v+\phi ^{0}+i\pi ^{0}\right)%
\end{array}%
\right) ,\hspace{1.5cm}\Delta =\left( 
\begin{array}{cc}
\frac{\Delta ^{+}}{\sqrt{2}} & \Delta ^{++} \\ 
v_{\Delta }+\Delta _{1}^{0}+i\Delta _{2}^{0} & -\frac{\Delta ^{+}}{\sqrt{2}}%
\end{array}%
\right) ,\hspace{1.5cm}S=v_{S}+\frac{1}{\sqrt{2}}\left( S_{R}+iS_{I}\right) .
\end{equation}
%
%
Since all singlet scalars acquire vacuum expectation values (VEVs) at a scale much larger than the electroweak symmetry breaking scale, they are very heavy and thus the mixing angles of these scalar singlets with the scalar fields $\phi$, $\Delta$ and $S$ are very suppressed by ratio of their VEVs, which is a consequence of the method of
recursive expansion proposed in Ref. \cite{Grimus:2000vj}. Consequently, we can neglect the mixing angles between these scalar fields, and at energies scales $\lesssim\mathcal{O}(1)$ TeV, it is enough to consider the low energy scalar potential of the model, 
studied in detail in Ref.~\cite{Ma:2017xxj}, where it was demonstrated that the Majoron $\eta_{I}$ resulting from the spontaneous breaking of the lepton number symmetry is given by the following linear combination
\begin{equation}
\eta_I=\frac{v_S\left(v^2+4v^2_{\Delta}\right)S_I+2v^2_{\Delta}v\pi^0-v^2v_{\Delta}\Delta^0_2}{\sqrt{v^2_S\left(v^2+4v^2_{\Delta}\right)^2+4v^2v^4_{\Delta}+v^4v^2_{\Delta}}}  
\end{equation}
In order to be consistent with 
the invisible $Z$-decay  and the precision electroweak  measurements one requires 
$v_{\Delta }<<v_{S}$  and $v_{\Delta }<<$ $v=246$ GeV, 
respectively.
%
We note that without $S$ our model at low energies would reduce to the triplet Majoron model \cite{Gelmini:1980re}, ruled out by the invisible $Z$-decay data.
%
%

We assume the following VEV pattern for the $\Delta \left( 27\right) $
triplet SM singlet scalars $\xi $, $\zeta $ and $\Phi$: 
\begin{equation}
\left\langle \xi \right\rangle =\frac{v_{\xi }}{\sqrt{3}}\left( 1,1,1\right)
,\hspace{1.5cm}\left\langle \zeta \right\rangle =\frac{v_{\zeta }}{\sqrt{2}}\left( 1,0,1\right),\hspace{1.5cm}\left\langle \Phi \right\rangle =v_{\Phi}\left(0,1,0\right) ,  \label{VEV}
\end{equation}%
which is a natural solution of the scalar potential minimization equations
for the whole region of parameter space, as shown in detail in 
Refs. \cite{Ivanov:2014doa,deMedeirosVarzielas:2017glw}

With the above particle content, the following Yukawa terms arise: 
\begin{eqnarray}
\tciLaplace _{Y} &=&y_{33}^{\left( U\right) }\overline{q}_{3L}\widetilde{%
\phi }u_{3R}+y_{23}^{\left( U\right) }\overline{q}_{2L}\widetilde{\phi }%
u_{3R}\frac{\sigma ^{2}}{\Lambda ^{2}}+y_{22}^{\left( U\right) }\overline{q}%
_{2L}\widetilde{\phi }u_{2R}\frac{\tau^{4}}{\Lambda ^{4}}+y_{11}^{\left(
U\right) }\overline{q}_{1L}\widetilde{\phi }u_{1R}\frac{\sigma ^{8}}{\Lambda
^{8}}  \notag \\
&&+y_{33}^{\left( D\right) }\overline{q}_{3L}\phi d_{3R}\frac{\eta ^{3}}{%
\Lambda ^{3}}+y_{13}^{\left( D\right) }\overline{q}_{1L}\phi d_{3R}\frac{%
\sigma \rho \tau ^{4}}{\Lambda ^{6}}+y_{22}^{\left( D\right) }\overline{q}%
_{2L}\phi d_{2R}\frac{\chi _{2}^{4}\sigma }{\Lambda ^{5}}+y_{12}^{\left(
D\right) }\overline{q}_{1L}\phi d_{2R}\frac{\chi _{1}^{4}\sigma ^{2}}{%
\Lambda ^{6}}+y_{11}^{\left( D\right) }\overline{q}_{1L}\phi d_{1R}\frac{%
\sigma ^{8}}{\Lambda ^{8}}  \notag \\
&&+y_{1}^{\left( l\right) }\left( \overline{l}_{L}\phi \xi \right) _{\mathbf{%
\mathbf{1}_{0\mathbf{,0}}}}l_{1R}\frac{\eta ^{7}\varphi _{1}}{\Lambda ^{9}}%
+y_{2}^{\left( l\right) }\left( \overline{l}_{L}\phi \xi \right) _{\mathbf{1}%
_{\mathbf{2,0}}}l_{2R}\frac{\eta ^{4}}{\Lambda ^{5}}+y_{3}^{\left( l\right)
}\left( \overline{l}_{L}\phi \xi \right) _{\mathbf{\mathbf{1}_{\mathbf{1,0}}}%
}l_{3R}\frac{\eta \varphi _{1}^{\ast }}{\Lambda ^{3}}+y_{4}^{\left( l\right)
}\left( \overline{l}_{L}\phi \xi \right) _{\mathbf{\mathbf{1}_{1\mathbf{,0}}}%
}l_{1R}\frac{\eta ^{7}\varphi _{2}^{\ast 2}}{\Lambda ^{10}}  \notag \\
&&+y_{5}^{\left( l\right) }\left( \overline{l}_{L}\phi \xi \right) _{\mathbf{%
\mathbf{1}_{0\mathbf{,0}}}}l_{3R}\frac{\eta \varphi _{2}^{2}}{\Lambda ^{4}}%
+y_{1}^{\left( \nu \right) }\left( \overline{l_{L}^{C}}i\sigma _{2}\Delta
l_{L}\right) _{\mathbf{3}_{S_{1}}}\frac{\zeta\eta^{8}}{\Lambda^{9} }%
+y_{2}^{\left( \nu \right) }\left( \overline{l_{L}^{C}}i\sigma _{2}\Delta
l_{L}\right) _{\mathbf{3}_{S_{2}}}\frac{\zeta\eta^{8}}{\Lambda^{9} } \notag\\
&&+y_{3}^{\left( \nu \right) }\left( \overline{l_{L}^{C}}i\sigma _{2}\Delta
l_{L}\right) _{\mathbf{3}_{S_{1}}}\frac{\Phi\eta^{8}}{\Lambda^{9} }%
+y_{4}^{\left( \nu \right) }\left( \overline{l_{L}^{C}}i\sigma _{2}\Delta
l_{L}\right) _{\mathbf{3}_{S_{2}}}\frac{\Phi\eta^{8}}{\Lambda^{9} }+h.c,
\label{Ly}
\end{eqnarray}%
where the dimensionless couplings 
are presumably $\mathcal{O}(1)$
parameters. Furthermore, as it will be shown in Sect. \ref{quarksector}, the
quark assignments under the different group factors of our model will give
rise to SM quark mass textures where the Cabbibo mixing as well as the
mixing in the 1-3 plane emerges from the down type quark sector, whereas the
up type quark sector generates the quark mixing angle in the 2-3 plane. 

In a generic scenario the Yukawa couplings are complex. However, not all of them are 
physical. Some phases can be rotated away by the phase rotation of the quark and lepton fields. The conditions for the rotation away of the Yukawa phases by the redefinition of the phases $\alpha_{f}$ of the fermion fields are:
%
\begin{eqnarray}
\label{PhaseconditionU}
\arg \left( y_{33}^{\left( U\right) }\right) -\alpha _{q_{3L}}+\alpha
_{u_{3R}} =0, &&  
\arg \left( y_{23}^{\left( U\right) }\right) -\alpha _{q_{2L}}+\alpha
_{u_{3R}} =0,  
\\ 
\notag
\arg \left( y_{22}^{\left( U\right) }\right) -\alpha _{q_{2L}}+\alpha
_{u_{2R}} =0,  &&
\arg \left( y_{11}^{\left( U\right) }\right) -\alpha _{q_{1L}}+\alpha
_{u_{1R}} = 0,  \\
\label{PhaseconditionD}
\arg \left( y_{33}^{\left( D\right) }\right) -\alpha _{q_{3L}}+\alpha
_{d_{3R}} =0, &&
\arg \left( y_{13}^{\left( D\right) }\right) -\alpha _{q_{1L}}+\alpha
_{d_{3R}} =0,  \notag \\
\arg \left( y_{22}^{\left( D\right) }\right) -\alpha _{q_{2L}}+\alpha
_{d_{2R}} =0,  &&
\arg \left( y_{12}^{\left( D\right) }\right) -\alpha _{q_{1L}}+\alpha
_{d_{2R}} = 0,  \notag \\
\notag
\arg \left( y_{11}^{\left( D\right) }\right) -\alpha _{q_{1L}}+\alpha_{d_{1R}} =0, && 
\end{eqnarray}%
\begin{eqnarray}
\label{PhaseconditionL}
\arg \left( y_{1}^{\left( l\right) }\right) -\alpha _{l_{L}}+\alpha
_{l_{1R}} =0,  &&
\arg \left( y_{2}^{\left( l\right) }\right) -\alpha _{l_{L}}+\alpha
_{l_{2R}} =0,  \notag \\
\arg \left( y_{3}^{\left( l\right) }\right) -\alpha _{l_{L}}+\alpha
_{l_{3R}} =0,  &&
\arg \left( y_{4}^{\left( l\right) }\right) -\alpha _{l_{L}}+\alpha
_{l_{1R}} =0,  \notag \\
\arg \left( y_{5}^{\left( l\right) }\right) -\alpha _{l_{L}}+\alpha
_{l_{3R}} =0,&&  
\\
\label{Phaseconditionnu}
\arg \left( y_{1,2,3,4}^{\left( \nu \right) }\right)\hspace{4mm} + 2\alpha _{l_{L}}  = 0.&&
\label{Phaseconditionnu}
\end{eqnarray}
As seen, in the quark sector all the  Yukawa phases can be rotated away. Thus, all the Yukawa coupling of the quark sector can be set real. 
In the sector of the charged leptons there is one physical phase. This can be the phase of one of the Yukawa couplings: $y^{(l)}_{1,3,4,5}$, while $y^{(l)}_{2}$ is always real. In the neutrino sector all the Yukawa phases are physical. 
%
Consequently, the observed CP violation in the quark sector will arise from complex vacuum expectation values of the gauge singlet scalars charged under the discrete symmetries of the model. Thus, the spontaneous breaking of the discrete symmetries of our model, gives rise to the observed CP violation in the quark sector. This mechanism of generating CP violation in the fermion sector from the spontaneous breaking of the discrete groups is called Geometrical CP violation and has been implemented in other models, such as, for example, in Refs.~ \cite{Branco:1983tn,Chen:2011tj,Bhattacharyya:2012pi,Girardi:2013sza,Varzielas:2013sla,Chen:2014tpa,Branco:2015hea,CarcamoHernandez:2019eme}. A concise review of group theoretical origin of CP violation is provided in Ref.~\cite{Chen:2019iup}. The geometrical CP violation can have implications for leptogenesis, see, for example, Ref.~\cite{Chen:2011tj}. 
%
%

Let us comment on the role of each discrete group factor of our model.
We were searching for  such discrete groups, which would allow us at minimal cost
to reach
viable textures for the fermion sector, consistent with the
observed pattern of fermion masses and mixings. We found that 
$\Delta \left( 27\right) \times Z_{16}\times Z_{24}$ is a good candidate, since 
%
the $Z_{16}$ and $Z_{24}$
symmetries give rise to the hierarchical structure of the charged fermion
mass matrices that yields the observed charged fermion mass and mixing
pattern. At the same time,
it is worth mentioning the properties of the $Z_{N}$ groups; in particular, the $Z_{16}$ symmetry is the smallest cyclic symmetry that allows us to build the Yukawa terms, $\overline{q}_{1L}\widetilde{\phi }u_{1R}\frac{
\sigma ^{8}}{\Lambda ^{8}}$ and $\overline{q}_{1L}\phi d_{1R}\frac{\sigma
^{8}}{\Lambda ^{8}}$, of dimension twelve from an $\frac{\sigma ^{8}}{\Lambda
^{8}}$ insertion on the $\overline{q}_{1L}\widetilde{\phi }u_{1R}$ and $\overline{q}_{1L}\phi d_{1R}$ operators. These are crucial to get the required $\lambda ^{8}$ suppression (where $\lambda =0.225$ is one of the Wolfenstein parameters) in the 11 entry of the quark mass matrices to naturally
explain the smallness of the up and down quark masses. In addition, it is noteworthy that
the small value of the up and down quark masses naturally arises from the aforementioned quark Yukawa terms of dimension 12. Along with this, we use the $Z_{24}$
discrete symmetry since it is the smallest cyclic symmetry that contains
both the $Z_{6}$ and $Z_{8}$ symmetries that glue $\eta ^{3}$ and $\tau ^{4}$
with $d_{3R}$, respectively, which is crucial for generating the right value
for the bottom quark mass and for the quark mixing angle in the $1$-$3$
plane, without tuning the corresponding Yukawa couplings.

Besides that, as the hierarchy among charged fermion masses and CKM parameters emerges from the spontaneous breaking of the $\Delta \left( 27\right) \times
Z_{16}\times Z_{24}$ discrete group, 
we set the magnitudes of the VEVs of the SM singlet
scalar fields $\sigma $,$\ \rho $,$\ \eta $,$\ \tau $,$\ \chi _{1}$, $\chi
_{2}$,$\ \varphi _{1}$, $\varphi _{2}$, $\xi _{i}$, $\zeta _{i}$ ($i=1,2,3$%
)\ with respect to the Wolfenstein parameter $\lambda =0.225$ and the model
cutoff $\Lambda $, as follows: 
\begin{equation}
|v_{\Phi}|\sim |v_{\zeta }|\sim |v_{\xi }|\sim |v_{\sigma }|\sim |v_{\rho }|\sim |v_{\eta }|\sim |v_{\tau
}|\sim |v_{\chi _{1}}|\sim |v_{\chi _{2}}|\sim |v_{\varphi _{1}}|\sim |v_{\varphi
_{2}}|\sim |v_{\xi }|\sim \lambda \Lambda .  \label{VEV}
\end{equation}
It is straightforward to show that the aforementioned assumption is consistent with the minimization 
conditions of the scalar potential (see, for instance, Refs.~\cite{Vien:2016tmh,CarcamoHernandez:2017owh}). To show that, it is sufficient to consider the quartic scalar couplings of order unity and the mass coefficients of the bilinear mass terms of the same order of magnitude. 

In the standard basis, the mass terms are given by 
\begin{equation}
\tciLaplace _{Y}=\bar{d}_{L}\mathbf{M}_{d}d_{R}+\bar{u}_{L}\mathbf{M}_{u}u_{R}+
\bar{l}_{L}\mathbf{M}_{l}l_{R}+\frac{1}{2}\bar{\nu}_{L}\mathbf{M}_{\nu }
(\nu_{L})^{C}+h.c,  \label{basis}
\end{equation}
where the mass matrices $\mathbf{M}_{d,u,l,\nu}$ are determined by the Yukawa couplings (\ref{Ly}) and will be studied in what follows.

\section{Quark masses and mixings}
\label{quarksector} 
The quark Yukawa interactions of Eq. (\ref{Ly}) and the
relation given by Eq. (\ref{VEV}), give rise to the following SM quark mass matrices
are:
\begin{equation}
\mathbf{M}_{u}
=\left( 
\begin{array}{ccc}
a_{1}^{\left( U\right) }\lambda ^{8} & 0 & 0 \\ 
0 & a_{2}^{\left( U\right) }\lambda ^{4} & a_{4}^{\left( U\right) }\lambda
^{2} \\ 
0 & 0 & a_{3}^{\left( U\right) }%
\end{array}%
\right) \frac{v}{\sqrt{2}},\hspace{1.5cm}
%
\mathbf{M}_{d}
=\left( 
\begin{array}{ccc}
a_{1}^{\left( D\right) }\lambda ^{8} & a_{4}^{\left( D\right) }\lambda ^{6}
& a_{5}^{\left( D\right) }\lambda ^{6} \\ 
0 & a_{2}^{\left( D\right) }\lambda ^{5} & 0 \\ 
0 & 0 & a_{3}^{\left( D\right) }\lambda ^{3}%
\end{array}%
\right) \frac{v}{\sqrt{2}},  \label{Quarktextures}
\end{equation}
where $\lambda =0.225$ is one of the Wolfenstein parameters, $v=246$ GeV the
scale of electroweak symmetry breaking and $a_{i}^{\left( U\right) }$ ($%
i=1,2,3,4$) and $a_{j}^{\left( D\right) }$ ($j=1,2,3,4$) are $\mathcal{O}(1)$
parameters. 
From the SM quark mass textures given above, it follows that the
Cabbibo mixing as well as the mixing in the 1-3 plane emerges from the down
type quark sector, whereas the up type quark sector generates the quark
mixing angle in the 2-3 plane. 

The hierarchical structure of the quark mass matrices (\ref{Quarktextures}) in the form of different powers of
 $\lambda$-parameter is a consequence of the symmetries of our model. If this hierarchical structure correctly reproduces the experimentally observed quark mass hierarchy 
we expect that all the dimensionless parameters are
$a^{(U,D)}_{i}\sim 1$.  As we will see, the values of these parameters, obtained from a fit to the quark masses and mixing angles, are compatible with this expectation. 
This observation, in particular, justifies the simplifying assumption $a_{4}^{(D)}=a_{1}^{(D)}$
allowing analytical diagonalization of the quark mass matrices. We do not use this simplification in our numerical analysis, but adopt it in order to derive certain analytical formulas, which help reveling some properties of our model.
With the simplified assumption $a_{4}^{(D)}=a_{1}^{(D)}$ we can write the quark mass matrices  (\ref{Quarktextures}) in the form
\begin{equation}
\label{eq:mass-matrices-quark-1}
\mathbf{M}_{u}
=\left( 
\begin{array}{ccc}
A_{U} & 0 & 0 \\ 
0 & B_{U} & C_{U} \\ 
0 & 0 & F_{U}
\end{array}
\right) , \hspace{1.5cm}
\mathbf{M}_{d}=\left( 
\begin{array}{ccc}
A_{D}\lambda ^{2} & A_{D} & F_{D} \\ 
0 & B_{D} & 0 \\ 
0 & 0 & C_{D}%
\end{array}%
\right) ,
%
\end{equation}%
with 
\begin{eqnarray}
\label{eq:ABCD-1} 
A_{D} &=&a_{1}^{\left( D\right) }\lambda ^{6}\frac{v}{\sqrt{2}},\quad
B_{D}=a_{2}^{\left( D\right) }\lambda ^{5}\frac{v}{\sqrt{2}},\quad
C_{D}=a_{3}^{\left( D\right) }\lambda ^{3}\frac{v}{\sqrt{2}},\quad
F_{D}=a_{5}^{\left( D\right) }\lambda ^{6}\frac{v}{\sqrt{2}};  \notag \\
A_{U} &=&a_{1}^{\left( U\right) }\lambda ^{8}\frac{v}{\sqrt{2}},\quad
B_{U}=a_{2}^{\left( U\right) }\lambda ^{4}\frac{v}{\sqrt{2}},\quad
C_{U}=a_{4}^{\left( U\right) }\lambda ^{2}\frac{v}{\sqrt{2}},\quad
F_{U}=a_{3}^{\left( U\right) }\frac{v}{\sqrt{2}}.
\end{eqnarray}%
In general, these mass matrices are complex and these can be diagonalized by the unitary
matrices ${\mathbf{U}_{u(L, R)}}$ and ${\mathbf{U}_{d(L, R)}}$ such that
\begin{equation}
\hat{\mathbf{M}}_{d}=\text{diag}\left(m_{d}, m_{s}, m_{b}\right)= \mathbf{U}%
^{\dagger}_{d L}\mathbf{M}_{d}\mathbf{U}_{d R},\hspace{1.5cm} \hat{\mathbf{M}%
}_{u}=\text{diag}\left(m_{u}, m_{c}, m_{t}\right)= \mathbf{U}^{\dagger}_{u L}%
\mathbf{M}_{u}\mathbf{U}_{u R}.
\end{equation}
In order to find the CKM matrix defined as
\begin{eqnarray}\label{eq:CKM-def-1}
\mathbf{V}^{CKM}=\mathbf{U}_{uL}^{\dagger }\mathbf{U}_{dL}
\end{eqnarray}
we proceed in the standard way and consider the matrices
%
$\hat{\mathbf{M}}_{d}\hat{
\mathbf{M}}^{\dagger}_{d}= \mathbf{U}^{\dagger}_{d L}\mathbf{M}_{d}\mathbf{M}%
^{\dagger}_{d}\mathbf{U}_{d L}$ and $\hat{\mathbf{M}}_{u}\hat{\mathbf{M}}%
^{\dagger}_{u}= \mathbf{U}^{\dagger}_{u L}\mathbf{M}_{u}\mathbf{M}%
^{\dagger}_{u}\mathbf{U}_{u L}$, 
{which are useful for derivation of} ${U}_{uL}$ and ${U}_{dL}$.
As one can be verified, the associated CP phases to the hermitian matrix $\mathbf{M}_{d}\mathbf{M}%
^{\dagger}_{d}$ can be factorized as $\mathbf{M}_{d}\mathbf{M}%
^{\dagger}_{d}=\mathbf {P}_{d} \mathbf{m}_{d}\mathbf{m}%
^{\dagger}_{d} \mathbf{P}%
^{\dagger}_{d} $ where $\mathbf{m}_{d}\mathbf{m}%
^{\dagger}_{d}$ is a real symmetric matrix and $\mathbf{P}_{d}= \text{Diag}\left(1, e^{-i\alpha_{D_{1}}},e^{-i\alpha_{D_{2}}}\right)$ where 

\begin{eqnarray}
\quad \alpha_{D_{1}}=\alpha_{A_{D}}-\alpha_{B_{D}}\quad \textrm{with}\quad \alpha_{A_{D}}= arg\left(A_{D}\right)\quad \textrm{and}\quad \alpha_{B_{D}}= arg\left(B_{D}\right)\nonumber\\
\quad \alpha_{D_{2}}=\alpha_{F_{D}}-\alpha_{C_{D}}\quad \textrm{with}\quad \alpha_{F_{D}}= arg\left(F_{D}\right)\quad \textrm{and}\quad \alpha_{C_{D}}= arg\left(C_{D}\right).
\end{eqnarray}

As usual we can write
$\mathbf{U}_{d L}=\mathbf{P}_{d} 
\mathbf{O}^{d}$. 
%
Explicitly, the real orthogonal matrix $\mathbf{O}^{d}$ is given by
\begin{equation}
\label{eq:O-quark-1}
\mathbf{O}^{d}=
\left(
\begin{array}{ccc}
-\sqrt{\frac{\left(\left\vert C_{D} \right\vert^{2}-\left\vert B_{D} \right\vert^{2}\right)
\left( \left\vert C_{D} \right\vert^{2}-m^{2}_{d}\right)\left(\left\vert B_{D} \right\vert^{2} -m^{2}_{d}\right)}{\mathcal{D}%
_{1}}} & \sqrt{\frac{\left(\left\vert C_{D} \right\vert^{2}-\left\vert B_{D} \right\vert^{2}\right)%
\left(\left\vert C_{D} \right\vert^{2}-m^{2}_{s}\right)\left(m^{2}_{s}-\left\vert B_{D} \right\vert^{2}\right)}{\mathcal{D}%
_{2}}} & \sqrt{\frac{\left(\left\vert C_{D} \right\vert^{2}-\left\vert B_{D} \right\vert^{2}\right)%
\left(m^{2}_{b}-\left\vert C_{D} \right\vert^{2}\right)\left(m^{2}_{b}-\left\vert B_{D} \right\vert^{2}\right)}{\mathcal{D}%
_{3}}} \\ 
\sqrt{\frac{\left(\left\vert C_{D} \right\vert^{2}-m^{2}_{d}\right)
\left(m^{2}_{b}-\left\vert B_{D} \right\vert^{2} \right)\left(m^{2}_{s}-\left\vert B_{D} \right\vert^{2}\right)}{\mathcal{D}%
_{1}}} & \sqrt{\frac{\left(\left\vert C_{D} \right\vert^{2}-m^{2}_{s}\right)%
\left(m^{2}_{b}-\left\vert B_{D} \right\vert^{2} \right)\left(\left\vert B_{D} \right\vert^{2} -m^{2}_{d}\right)}{\mathcal{D}%
_{2}}} & \sqrt{\frac{\left(m^{2}_{b}-\left\vert C_{D} \right\vert^{2}\right)%
\left(m^{2}_{s}-\left\vert B_{D} \right\vert^{2}\right)\left(\left\vert B_{D} \right\vert^{2}-m^{2}_{d}\right)}{\mathcal{D}%
_{3}}} \\ 
\sqrt{\frac{\left(\left\vert C_{D} \right\vert^{2}-m^{2}_{s}\right)
\left(m^{2}_{b}-\left\vert C_{D} \right\vert^{2}\right)\left(\left\vert B_{D} \right\vert^{2}-m^{2}_{d}\right)}{\mathcal{D}%
_{1}}} & -\sqrt{\frac{\left(\left\vert C_{D} \right\vert^{2}-m^{2}_{d}\right)%
\left(m^{2}_{b}-\left\vert C_{D} \right\vert^{2} \right)\left(m^{2}_{s}-\left\vert B_{D} \right\vert^{2} \right)}{\mathcal{D}%
_{2}}} & \sqrt{\frac{\left(\left\vert C_{D} \right\vert^{2}-m^{2}_{s}\right)%
\left(\left\vert C_{D} \right\vert^{2} -m^{2}_{d}\right)\left(m^{2}_{b}-\left\vert B_{D} \right\vert^{2} \right)}{\mathcal{D}%
_{3}}}
\end{array}
\right)
\end{equation}
with 
\begin{eqnarray}
\mathcal{D}_{1}&=&\left(\left\vert C_{D} \right\vert^{2}-\left\vert B_{D} \right\vert^{2}\right)\left(m^{2}_{b}-m^{2}_{d}
\right)\left(m^{2}_{s}-m^{2}_{d}\right),  \notag \\
\mathcal{D}_{2}&=&\left(\left\vert C_{D} \right\vert^{2}-\left\vert B_{D} \right\vert^{2}\right)
\left(m^{2}_{b}-m^{2}_{s}
\right)\left(m^{2}_{s}-m^{2}_{d}\right),  \notag \\
\mathcal{D}_{3}&=&\left(\left\vert C_{D} \right\vert^{2}-\left\vert B_{D} \right\vert^{2}\right)
\left(m^{2}_{b}-m^{2}_{s}
\right)\left(m^{2}_{b}-m^{2}_{d}\right).
\end{eqnarray}
{As seen from (\ref{eq:O-quark-1}), }there is a condition on the parameters  
$m_{b}>\left\vert C_{D} \right\vert>m_{s}>\left\vert B_{D} \right\vert >m_{d}$.

In the up quark sector, analogously to the down quark sector,
we obtain that $\mathbf{M}_{u}\mathbf{M}%
^{\dagger}_{u}=\mathbf {P}_{u} \mathbf{m}_{u}\mathbf{m}%
^{\dagger}_{u} \mathbf{P}%
^{\dagger}_{u} $ where $\mathbf{m}_{u}\mathbf{m}%
^{\dagger}_{u}$ is a real symmetric matrix and $\mathbf{P}_{u}= \text{Diag}\left(1, 1,e^{-i\alpha_{U}}\right)$ where 
\begin{equation}
\alpha_{U}=\alpha_{C_{U}}-\alpha_{F_{U}}\qquad \textrm{with}\quad \alpha_{C_{U}}= arg\left(C_{U}\right)\quad \textrm{and}\quad \alpha_{F_{U}}= arg\left( F_{U}\right).
\end{equation}

Then, we choose appropriately $\mathbf{U}_{u L}=\mathbf{P}_{u} \mathbf{O}_{u}$ where

%
\begin{equation}
\mathbf{O}_{u }=\left(
\begin{array}{ccc}
1 & 0 & 0 \\ 
0 & \cos{\theta_{u}} & \sin{\theta_{u}} \\ 
0 & -\sin{\theta_{u}} & \cos{\theta_{u}}%
\end{array}%
\right),\quad \cos{\theta_{u}}=\frac{1}{\sqrt{2}}\sqrt{\frac{%
m^{2}_{t}-m^{2}_{c}-\left\vert C_{U}\right\vert^{
2}+R_{u}}{m^{2}_{t}-m^{2}_{c}}},\quad \sin{%
\theta_{u}}=\frac{1}{\sqrt{2}}\sqrt{\frac{m^{2}_{t}-m^{2}_{c}+
\left\vert C_{U}\right\vert^{2}-R_{u}%
}{m^{2}_{t}-m^{2}_{c}}}
\end{equation}
with $R_{u}=\sqrt{(m^{2}_{t}+m^{2}_{c}-\left\vert C_{U}\right\vert^{
2})^{2}-4m^{2}_{t}m^{2}_{c}}$, and 
the condition $m_{t}>\left\vert C_{U}\right\vert>m_{c}$ to be satisfied.

Then, we obtain the CKM mixing matrix (\ref{eq:CKM-def-1}) in the form
\begin{equation}
\mathbf{V}^{CKM}=\left( 
\begin{array}{ccc}
O_{11}^{d} & O_{12}^{d} & O_{13}^{d} \\ 
O_{21}^{d}\cos {\theta _{u}}e^{-i\alpha_{D_{1}}}-O_{31}^{d}\sin {\theta _{u}}e^{-i\bar{\alpha}
_{D_{2}}} & O_{22}^{d}\cos {\theta _{u}}e^{-i\alpha_{D_{1}}}-O_{32}^{d}\sin {\theta _{u}}%
e^{-i\bar{\alpha} _{D_{2}}} & O_{23}^{d}\cos {\theta _{u}}e^{-i\alpha_{D_{1}}}-O_{33}^{d}\sin {\theta
_{u}}e^{-i\bar{\alpha} _{D_{2}}} \\ 
O_{21}^{d}\sin {\theta _{u}}e^{-i\alpha_{D_{1}}}+O_{31}^{d}\cos {\theta _{u}}e^{-i\bar{\alpha}
_{D_{2}}} & O_{22}^{d}\sin {\theta _{u}}e^{-i\alpha_{D_{1}}}+O_{32}^{d}\cos {\theta _{u}}%
e^{-i\bar{\alpha} _{D_{2}}} & O_{23}^{d}\sin {\theta _{u}}e^{-i\alpha_{D_{1}}}+O_{33}^{d}\cos {\theta
_{u}}e^{-i\bar{\alpha} _{D_{2}}}%
\end{array}%
\right),
\end{equation}
where $\bar{\alpha} _{D_{2}}=\alpha_{D_{2}}-\alpha_{U}$. Note that this mixing matrix depends on four free parameters $\left\vert C_{U}\right\vert$, $\left\vert C_{D}\right\vert$, $\left\vert B_{D}\right\vert$
and one effective CP violating phase to be adjusted in order to reproduce the CKM matrix elements.
However, as one can verify, the correct value of the Cabbibo angle is obtained with the following
values of  the model parameters $\left\vert C_{U}\right\vert^{2} =m_{c}^{2}+m_{t}m_{c}$, $%
\left\vert C_{D}\right\vert^{2}=m_{b}^{2}$ and $ \left\vert B_{D}\right\vert^{2}=m_{s}^{2}-m_{s}m_{d}$. Then, we have 
\begin{equation}
V^{CKM}_{us}\approx \sqrt{\frac{m_{d}}{m_{s}}}\left( 1+\frac{1}{2}\frac{m_{d}^{2}}{
m_{s}^{2}}\right) ,\qquad V^{CKM}_{cb}\approx -\sqrt{\frac{m_{c}}{m_{t}}},\qquad
V^{CKM}_{td}\approx \sqrt{\frac{m_{d}}{m_{s}}}\sqrt{\frac{m_{c}}{m_{t}}}.
\end{equation}%
These approximate relations, valid in our model, are in agreement with the well known 
extended Gatto-Sartori-Tonin relations
\cite{Gatto:1968ss,Cabibbo:1968vn,Oakes:1969vm,Fritzsch:1977za,Fritzsch:1977vd,Fritzsch:1979zq,Fritzsch:1999ee,Fritzsch:1999rb,Gupta:2013yha,Saldana-Salazar:2015raa,Saldana-Salazar:2018jes,Rahat:2018sgs}.

After these notes on some properties of our model derived from the explicit analytical expression for the quark spectrum and CKM mixing matrix we carry out a numerical analysis without imposing $a_{4}^{(D)}=a_{1}^{(D)}$. We find that the experimental values for the physical quark mass spectrum~\cite{Bora:2012tx,Xing:2007fb}, mixing angles and CP violating phase~\cite{Patrignani:2016xqp} can be reproduced for the following benchmark point:
%
\begin{eqnarray}
a_{1}^{\left( U\right) } &\simeq &1.252,\hspace{1cm}a_{2}^{\left( U\right)}\simeq 1.415,\hspace{1cm}a_{3}^{\left(U\right) }\simeq 0.989,\hspace{1cm}a_{4}^{\left( U\right) }\simeq 0.802,\hspace{1cm}a_{1}^{\left(D\right) }\simeq 0.579,  \notag \\
a_{2}^{\left( D\right) } &\simeq &0.570,\hspace{1cm}a_{3}^{\left( D\right)
}\simeq 1.416,\hspace{1cm}a_{4}^{\left( D\right) }\simeq 0.583,\hspace{1cm}
a_{5}^{\left( D\right) }\simeq 0.163+0.403i.
\label{fit-q}
\end{eqnarray}
Naturally, ten free parameters fit perfectly ten observables, but what is important is that the absolute values of all the  parameters are $|a_{i}^{(A)}| \sim 1$. This means that the hierarchies existing in the quark spectrum and mixing is reproduced by the symmetries of the model resulting in the particular texture of the quark mass matrices 
(\ref{Quarktextures}) without the need of manual introduction of a hierarchy in the free parameters $a_{i}^{(A)}$. We only mildly tune these parameters to perfectly reproduce the quark masses and mixing.

The result of the fit in Eq.~(\ref{fit-q}) suggests several simplified benchmark 
scenarios: 
\begin{eqnarray}\label{eq:S-5}
\mbox{S-5 (5 parameters):}&& a_{4}^{\left( D\right) }=a_{1}^{\left(
D\right) },\ \ \ a_{1}^{\left( U\right) }=a_{3}^{\left( U\right) }=1,\ \ \
a_{3}^{\left( D\right) }=a_{2}^{\left( U\right) }.\\
\nonumber
\mbox{Best-fit values:}&&
a_{2}^{\left( U\right) }\simeq 1.43,\ \ \ a_{4}^{\left( U\right)
}\simeq 0.80, \ \ \ a_{1}^{\left( D\right) }\simeq 0.58, \ \ \ 
a_{2}^{\left( D\right) }\simeq 0.57, \ \ \ \left\vert a_{5}^{\left(
D\right) }\right\vert \simeq 0.44.\\
\label{eq:S-4}
\mbox{S-4 (4 parameters):}&& a_{4}^{\left( D\right) }=a_{1}^{\left(
D\right) }=a_{2}^{(D)},\ \ \ a_{1}^{\left( U\right) }=a_{3}^{\left( U\right)
}=1,\ \ \ a_{3}^{\left( D\right) }=a_{2}^{\left( U\right) }.\\
\nonumber
\mbox{Best-fit values:}&&
a_{2}^{\left( U\right) }\simeq 1.42,\hspace{1cm}a_{4}^{\left( U\right)
}\simeq {0.81},\hspace{1cm}a_{1}^{\left( D\right) }\simeq {0.58},\hspace{1cm}%
\left\vert a_{5}^{\left( D\right) }\right\vert \simeq {0.43}.\\
\label{eq:S-3}
\mbox{S-3 (3 parameters):} && a_{4}^{\left( D\right) }=a_{1}^{\left(
D\right) }=a_{2}^{(D)},\ \ \ a_{1}^{\left( U\right) }=a_{3}^{\left( U\right)
}=a_{4}^{(U)}=1,\ \ \ a_{3}^{\left( D\right) }=a_{2}^{\left( U\right) }.\\
\nonumber
\mbox{Best-fit values:}&&
a_{2}^{\left( U\right) }\simeq {1.42},\hspace{1cm}a_{1}^{\left( D\right)
}\simeq {0.58},\hspace{1cm}\left\vert a_{5}^{\left( D\right) }\right\vert
\simeq {0.43}.\\
\label{eq:S-2}
\mbox{S-2 (2 parameters):} &&a_{4}^{\left( D\right) }=a_{1}^{\left(
D\right) }=a_{2}^{(D)},\ \ \ a_{1}^{\left( U\right) }=a_{3}^{\left( U\right)
}=a_{4}^{(U)}=a_{3}^{\left( D\right) }=a_{2}^{\left( U\right) }=1.\\
\nonumber
\mbox{Best-fit values:}&&
a_{1}^{\left( D\right) }\simeq {0.58},\hspace{1cm}\left\vert a_{5}^{\left(
D\right) }\right\vert \simeq {0.31}.
\end{eqnarray}

\begin{table}[tbh]
\begin{center}
\begin{tabular}{c|l|l|l|l|l}
\hline\hline
Observable & S-5& S-4 & S-3 & S-2& Experimental value \\ \hline
$m_{u}(MeV)$ & \quad $1.16$ & $1.16$ & $1.16$ & $1.16$ & \quad $%
1.45_{-0.45}^{+0.56}$ \\ \hline
$m_{c}(MeV)$ & \quad $641$ & $635$ & $634$ & $448$ & \quad $635\pm 86$ \\ 
\hline
$m_{t}(GeV)$ & \quad $174$ & $174$ & $174$ & $174$ & \quad $172.1\pm 0.6\pm
0.9$ \\ \hline
$m_{d}(MeV)$ & \quad $2.9$ & $2.9$ & $2.9$ & $2.9$ & \quad $2.9_{-0.4}^{+0.5}
$ \\ \hline
$m_{s}(MeV)$ & \quad $59.2$ & $60.1$ & $60.1$ & $60.1$ & \quad $%
57.7_{-15.7}^{+16.8}$ \\ \hline
$m_{b}(GeV)$ & \quad $2.85$ & $2.82$ & $2.82$ & $1.99$ & \quad $%
2.82_{-0.04}^{+0.09}$ \\ \hline
$\sin \theta _{12}$ & \quad $0.225$ & $0.220$ & $0.220$ & $0.220$ & \quad $%
0.225$ \\ \hline
$\sin \theta _{23}$ & \quad $0.0407$ & $0.0411$ & $0.0507$ & $0.0507$ & 
\quad $0.0412$ \\ \hline
$\sin \theta _{13}$ & \quad $0.00352$ & $0.00351$ & $0.00351$ & $0.00351$ & 
\quad $0.00351$ \\ \hline\hline
\end{tabular}%
\end{center}
\caption{Values of the CKM parameters and quark masses in our model compared with the experimental ones. The columns label S-P correspond to the different benchmark scenarios, where P stands for the number of free parameters. The quark masses are given at the $M_Z$ scale.}
\label{Tab-1}
\end{table}

As seen from Tab.~\ref{Tab-1} the 5- and 4-parameter scenarios  S-5 and S-4 fit perfectly the experimental data. Even the 2-parameter scenario S-2 does not contradict the data.
In view of this fact it becomes tempting to think  that the conditions  
(\ref{eq:S-5}),  (\ref{eq:S-4}), (\ref{eq:S-3}), (\ref{eq:S-2}), imposed as additional constraints, actually originate from some symmetry compatible with our model, but missed  in the present study. 
We will address this possibility elsewhere.

\section{Lepton masses and mixings}
\label{leptonsector} 
Analyzing the lepton sector of the model we adopt a simplifying benchmark scenario with particular assumption about the model parameters
\begin{equation}
y_{4}^{\left( l\right) }=y_{1}^{\left( l\right) },\hspace{1cm}y_{5}^{\left(
l\right) }=-y_{3}^{\left( l\right) },\hspace{1cm}v_{\varphi _{1}}=\lambda
\cos \theta _{l}\Lambda ,\hspace{1cm}v_{\varphi _{2}}=\sqrt{\lambda \sin
\theta _{l}}\Lambda ,\hspace{1cm}v_{\eta }=\lambda \Lambda,
\label{Benchmark}
\end{equation}
where $\sin \theta _{l}\sim \mathcal{O}(\lambda )$.  
This scenario is compatible with the VEV hierarchy (\ref{VEV}) and, as we will show, features a nice simple relation between the reactor mixing angle and the Wolfenstein parameter. 

In what follows we limit ourselves to this scenario considering the relations  (\ref{Benchmark}) as additional constraints on our model parameter space. 
On the other hand it is possible that the relations (\ref{Benchmark}) arise in our model framework as a consequence of some unrecognized symmetry or can be attributed to a particular ultraviolet completion of the model.

From the lepton Yukawa terms in
Eq. (\ref{Ly}) we find the charged lepton mass matrix $M_{l}$ defined in (\ref{basis}).  It can be written in the form
\begin{equation}
\mathbf{M}_{l}=R_{lL}S_{lL}{\rm Diag}\left( m_{e},m_{\mu },m_{\tau }\right) ,\hspace{0.5cm}%
R_{lL}=\frac{1}{\sqrt{3}}\left( 
\begin{array}{ccc}
1 & 1 & 1 \\ 
1 & \omega & \omega ^{2} \\ 
1 & \omega ^{2} & \omega
\end{array}
\right) ,\hspace{0.5cm}\omega =e^{\frac{2\pi i}{3}},\hspace{0.5cm}
S_{l}=\left( 
\begin{array}{ccc}
\cos \theta _{l} & 0 & -\sin \theta _{l} \\ 
0 & 1 & 0 \\ 
\sin \theta _{l} & 0 & \cos \theta _{l}
\end{array}
\right) ,
\end{equation}
where the charged lepton masses are given by: 
\begin{equation}
m_{e}=a_{1}^{\left( l\right) }\lambda ^{9}\frac{v}{\sqrt{2}},\hspace{1.5cm}%
m_{\mu }=a_{2}^{\left( l\right) }\lambda ^{5}\frac{v}{\sqrt{2}},\hspace{1.5cm%
}m_{\tau }=a_{3}^{\left( l\right) }\lambda ^{3}\frac{v}{\sqrt{2}}.
\label{leptonmasses}
\end{equation}%
Here $a_{i}^{\left( l\right) }$ ($i=1,2,3 $) are $\mathcal{O}(1)$ dimensionless parameters. 

In the standard basis (\ref*{basis}) 
the charged lepton mass matrix is diagonalized with 
$\mathbf{U}_{lL}=R_{lL}S_{l}$ so that 
$\mathbf{U}_{lL}^{\dagger }\mathbf{M}_{l} \mathbf{U}_{lL}=
{S}_{l}^{T}{R}_{lL}^{\dagger }{R}_{lL}
{S}_{l}\cdot {\rm Diag}(m_{e},m_{\mu },m_{\tau })={\rm Diag}(m_{e},m_{\mu },m_{\tau })$. 

The neutrino mass matrix, derived from the Yukawa terms in Eq.~(\ref{Ly}), is given by:
\begin{equation}
\mathbf{M}_{\nu }=%
\begin{pmatrix}
a_{\nu } & b_{\nu } & c_{\nu } \\ 
b_{\nu } & d_{\nu } & b_{\nu } \\ 
c_{\nu } & b_{\nu } & a_{\nu }%
\end{pmatrix}%
\end{equation}%
where
\begin{equation}
a_{\nu }=y_{1}^{\left( \nu \right) }\frac{v_{\zeta }v^8_{\eta }}{\sqrt{2}\Lambda^9}%
v_{\Delta },\hspace{0.7cm}b_{\nu }=y_{2}^{\left( \nu \right) }\frac{v_{\zeta }v^8_{\eta }%
}{\sqrt{2}\Lambda^9}v_{\Delta },\hspace{0.7cm}c_{\nu }=y_{4}^{\left( \nu
\right) }\frac{v_{\Phi}v^8_{\eta }}{\Lambda^9}v_{\Delta },\hspace{0.7cm}d_{\nu
}=y_{3}^{\left( \nu \right) }\frac{v_{\Phi}v^8_{\eta }}{\Lambda^9}v_{\Delta }.
\end{equation}
Diagonalization of neutrino mass matrix  
$\mathbf{U}_{\nu }^{\dagger }\mathbf{M}_{\nu }\mathbf{U}_{\nu
}^{\ast }=\mathbf{\hat{M}}=\text{diag}\left( m_{1},m_{2},m_{3}\right) $
is realized by 
$\mathbf{U}_{\nu }=\mathbf{U}_{\pi /4}\mathbf{u}_{\nu }$ so that 
\begin{equation}
\mathbf{\hat{M}}=\mathbf{u}_{\nu }^{\dagger }\mathbf{m}_{\nu }\mathbf{u}%
_{\nu }^{\ast },\hspace{0.5cm}\mathbf{m}_{\nu }=\left( 
\begin{array}{ccc}
a_{\nu }+c_{\nu } & \sqrt{2}b_{\nu } & 0 \\ 
\sqrt{2}b_{\nu } & d_{\nu } & 0 \\ 
0 & 0 & a_{\nu }-c_{\nu }%
\end{array}%
\right) ,\hspace{0.5cm}\mathbf{U}_{\pi /4}=\left( 
\begin{array}{ccc}
\frac{1}{\sqrt{2}} & 0 & -\frac{1}{\sqrt{2}} \\ 
0 & 1 & 0 \\ 
\frac{1}{\sqrt{2}} & 0 & \frac{1}{\sqrt{2}}%
\end{array}%
\right) ,
\end{equation}
where
\begin{equation}
\mathbf{u}_{\nu}= \left( 
\begin{array}{ccc}
\cos{\theta}_{\nu} & \sin{\theta}_{\nu} & 0 \\ 
-\sin{\theta}_{\nu} & \cos{\theta}_{\nu} & 0 \\ 
0 & 0 & 1%
\end{array}%
\right).
\end{equation}
with 
\begin{eqnarray}\label{eq:ThetaNu-1}
\tan{2\theta_{\nu}}&=&\frac{\sqrt{8}b_{\nu}}{d_{\nu}-a_{\nu}-c_{\nu}}
\end{eqnarray}
and the neutrino masses given by
\begin{eqnarray}\label{nmasses}
m_{1}&=&\left(a_{\nu}+c_{\nu}\right)\cos^{2}{\theta_{\nu}}+d_{\nu}\sin^{2}{%
\theta_{\nu}}-\sqrt{2}b_{\nu}\sin{2\theta_{\nu}},  \notag \\
m_{2}&=&\left(a_{\nu}+c_{\nu}\right)\sin^{2}{\theta_{\nu}}+d_{\nu}\cos^{2}{%
\theta_{\nu}}+\sqrt{2}b_{\nu}\sin{2\theta_{\nu}},  \notag \\
m_{3}&=&a_{\nu}-c_{\nu},  \notag \\
\end{eqnarray}
Considering the PMNS mixing matrix, defined as $\mathbf{V}=\mathbf{U}_{lL}^{\dagger}\mathbf{U}_{\nu }$, we find its matrix elements:
\begin{eqnarray}\label{pmnsen}
V_{11} &=&\sqrt{\frac{2}{3}}\cos {\theta _{l}}\cos {\theta _{\nu }}\left[ 1-%
\frac{\omega}{2}\tan {\theta _{l}}-\frac{\sqrt{2}}{2}\tan {\theta
_{\nu }}\left( 1+\omega\tan {\theta _{l}}\right) \right] ,  \notag
\\
V_{12} &=&\frac{1}{\sqrt{3}}\cos {\theta _{l}}\cos {\theta _{\nu }}\left[
1+\omega\tan {\theta _{l}}+\sqrt{2}\tan {\theta _{\nu }}\left( 1-%
\frac{\omega}{2}\tan {\theta _{l}}\right) \right] ,  \notag \\
V_{13} &=&-\frac{i\omega}{\sqrt{2}}\sin {\theta _{l}},  \notag \\
V_{21} &=&-\frac{\omega^2 }{\sqrt{6}}\cos {\theta _{\nu }}\left[ 1+\sqrt{2}%
\tan {\theta _{\nu }}\right] ,  \notag \\
V_{22} &=&\frac{\omega^2}{\sqrt{3}}\cos {\theta _{\nu }}\left[ 1-\frac{1}{ 
\sqrt{2}}\tan {\theta _{\nu }}\right] ,  \notag \\
V_{23} &=&\frac{i\omega^2}{\sqrt{2}},  \notag \\
V_{31} &=&-\frac{\omega}{\sqrt{6}}\cos {\theta _{l}}\cos {\theta
_{\nu }}\left[ 1+\sqrt{2}\tan {\theta _{\nu }}+\sqrt{2}\omega^2 \tan {\theta
_{l}}\left( \sqrt{2}-\tan {\theta _{\nu }}\right) \right] ,  \notag \\
V_{32} &=&\frac{\omega}{\sqrt{3}}\cos {\theta _{l}}\cos {\theta
_{\nu }}\left[ 1-\frac{1}{\sqrt{2}}\tan {\theta _{\nu }}-\omega^2 \tan {\theta
_{l}}\left( 1+\sqrt{2}\tan {\theta _{\nu }}\right) \right] ,  \notag \\
V_{33} &=&-\frac{i\omega }{\sqrt{2}}\cos {\theta _{l}}.
\end{eqnarray}
In the standard parametrization of the PMNS matrix we have the generic relations
\begin{equation}  \label{mixan}
\sin ^{2}{\theta _{13}}=\left\vert V_{13}\right\vert ^{2},\hspace{0.5cm}
\sin^{2}{\theta _{23}}=\frac{\left\vert V_{23}\right\vert ^{2}}{1-\left\vert
V_{13}\right\vert ^{2}},\hspace{0.5cm}\sin ^{2}{\theta _{12}}=
\frac{\left\vert V_{12}\right\vert ^{2}}{1-\left\vert V_{13}\right\vert ^{2}}.
\end{equation}
Examining Eqs.~(\ref{pmnsen}) and (\ref{mixan}) we get a nice relation between
the reactor and atmospheric angles:
\begin{eqnarray}\label{eq:13-23Relation}
&&\sin^{2}{\theta _{23}}=\frac{1/2}{1-\sin^{2}\theta_{13}},
\end{eqnarray}
which satisfies the experimental data in Tables~\ref{neutrinoDataNH},\ref{neutrinoDataIH}
within $\sim 1 \sigma$ both for normal hierarchy (NH) and inverted hierarchy (IH). 
As seen from Eqs.~(\ref{pmnsen}) and (\ref{mixan}) the model has only two free parameters  
$\theta_{l}$ and 
$\theta_{\nu}$ to fit four neutrino sector observables: three angles 
$\theta_{12}, \theta_{23}, \theta_{13}$ and the CP-violating phase $\delta_{CP}$. 

On the other hand, according to Eqs.~(\ref{nmasses}), (\ref{eq:ThetaNu-1}), there are three parameters 
$a_{\nu}, b_{\nu}, c_{\nu}$ to fit two observables $\Delta m^{2}_{21}$ and 
$\Delta m^{2}_{31/13}$, which means the model has no limitations on the neutrino mass squared
differences. Any of their experimental values can be reproduced exactly.  That is why in Tables~\ref{neutrinoDataNH},\ref{neutrinoDataIH} we do not show model values for these two observables.

Then we fit only the neutrino mixing angles $\theta_{ij}$ and the
CP-violating phase $\delta_{CP}$. The latter can be conveniently
extracted from the neutrino mixing matrix (\ref{pmnsen}) using the
Jarlskog invariant
\begin{equation}
\label{eq:JCP-1}
J_{CP}={\rm Im}\left[V_{23}V^{\ast}_{13}V_{12}V^{\ast}_{22}\right]
\end{equation}
and its equivalent definition \cite{Krastev:1988yu} in the standard parametrization 
\begin{equation}\label{JI1}
J_{CP}=\frac{1}{8}\sin{2\theta_{12}}\sin{2\theta_{23}}\sin{2\theta_{13}}\cos{\theta_{13}}\sin{\delta_{CP}}.
\end{equation}
Comparing  Eqs. (\ref{eq:JCP-1}) and (\ref{JI1}) as well as taking into account (\ref{pmnsen}), (\ref{mixan}) we find an expression for $\sin\delta_{CP}$ in terms of the model parameters 
$\theta_{l}$ and 
$\theta_{\nu}$. 
\begin{table}[tbh]
\begin{center}
\begin{tabular}{|c||c|c|c|c|}
\hline
\multirow{2}{*}{Observable} & \multirow{2}{*}{Model value} &  \multicolumn{3}{|c|}{Experimental value} \\ \cline{3-5}
 & & $1\sigma$ range & $2\sigma$ range & $3\sigma$ range \\ \hline\hline
$\Delta m_{21}^{2}$ [$10^{-5}$eV$^{2}$] (NH) & -- & $7.55_{-0.16}^{+0.20} $ & $7.20-7.94$ & $7.05-8.14$ \\ \hline
$\Delta m_{31}^{2}$ [$10^{-3}$eV$^{2}$] (NH) & -- & $2.50\pm 0.03$ & $2.44-2.57$ & $2.41-2.60$ \\ \hline
$\delta_{CP} $ [$^{\circ }$] (NH) & $233$ & $218_{-27}^{+38}$ & $182-315$ & $157-349$ \\ \hline
$\sin ^{2}\theta _{12}/10^{-1}$ (NH) & $3.2$ & $3.20_{-0.16}^{+0.20} $ & $2.89-3.59$
& $2.73-3.79$ \\ \hline
$\sin ^{2}\theta _{23}/10^{-1}$ (NH) & $5.11$ & $5.47_{-0.30}^{+0.20}$ & $4.67-5.83$
& $4.45-5.99$ \\ \hline
$\sin ^{2}\theta _{13}/10^{-2}$ (NH) & $2.170$ & $2.160_{-0.069}^{+0.083}$ & $%
2.03-2.34$ & $1.96-2.41$ \\ \hline
\end{tabular}%
\end{center}
\caption{The model values shown in the table correspond to the results of the best fit for the mixing angles and CP violating phase for normal hierarchy. The $1-3\sigma$ experimental ranges \protect\cite{deSalas:2017kay} are also shown for comparison.}
\label{neutrinoDataNH}
\end{table}
\begin{table}[tbh]
\begin{center}
\begin{tabular}{|c||c|c|c|c|}
\hline
\multirow{2}{*}{Observable} & \multirow{2}{*}{Model value} &  \multicolumn{3}{|c|}{Experimental value} \\ \cline{3-5}
 & & $1\sigma$ range & $2\sigma$ range & $3\sigma$ range \\ \hline\hline
$\Delta m_{21}^{2}$ [$10^{-5}$eV$^{2}$] (IH) & -- & $7.55_{-0.16}^{+0.20} $ & $7.20-7.94$ & $7.05-8.14$ \\ \hline
$\Delta m_{13}^{2}$ [$10^{-3}$eV$^{2}$] (IH) & -- & $2.42_{-0.04}^{+0.03} $ & $2.34-2.47$ & $2.31-2.51$ \\ \hline
$\delta_{CP} $ [$^{\circ }$] (IH) & 233 & $281_{-27}^{+23}$ & $229-328$ & $%
202-349$ \\ \hline
$\sin ^{2}\theta _{12}/10^{-1}$ (IH) & $3.20$ & $3.20_{-0.16}^{+0.20} $ & $2.89-3.59$
& $2.73-3.79$ \\ \hline
$\sin ^{2}\theta _{23}/10^{-1}$ (IH) & $5.11$ & $5.51_{-0.30}^{+0.18}$ & $4.91-5.84$
& $4.53-5.98$ \\ \hline
$\sin ^{2}\theta _{13}/10^{-2}$ (IH) & $2.223$ & $2.220_{-0.076}^{+0.074}$ & $%
2.07-2.36$ & $1.99-2.44$ \\ \hline
\end{tabular}
\end{center}
\caption{The model values shown in the table correspond to the results of the best fit for the mixing angles and CP violating phase for inverted hierarchy. The $1-3\sigma$ experimental ranges \protect\cite{deSalas:2017kay} are also shown for comparison.}\label{neutrinoDataIH}   
\end{table}

In Tables~\ref{neutrinoDataNH},\ref{neutrinoDataIH} we show the best-fit values of the mixing angles $\sin^{2}\theta_{ij}$ and the CP-violating phase $\delta_{CP}$ corresponding to the model parameters 
\begin{eqnarray}
&&\mbox{NH}: \theta_l\approx 12^{\circ }, \ \ \theta_{\nu} \approx  -65^{\circ }, \label{NH}\nonumber \\[0.12in]
&&\mbox{IH}: \theta_l\approx 12.2^{\circ }, \ \  \theta_{\nu} \approx  -65^{\circ }. \label{IH} 
\label{fitpointneutrino}
\end{eqnarray}
Varying the angles $\theta_{\nu}$, $\theta_{l}$ within $\sim 1\sigma$ range of the experimental values of $\sin^{2}\theta_{ij}$ we find that in the benchmark scenario  (\ref{Benchmark})  
our model predicts
\begin{eqnarray}\label{eq:Jarlskog-variation}
-3.0\times 10^{-2}\lesssim J_{CP}\lesssim -2.40\times 10^{-2}, \ \ \ 230^{\circ}\lesssim\delta_{CP}\lesssim 236^{\circ}
\end{eqnarray}
for both the normal and inverted neutrino mass spectrum.

The two Majorana phases, $\alpha$ and $\beta$, can be calculated through the invariants \cite{Nieves:1987pp, Bilenky:2001xq,Nieves:2001fc,AguilarSaavedra:2000vr}
\begin{equation}\label{IM1}
I_{1}= {\rm Im}\left[ V^{\ast}_{11}V_{12}\right],\qquad I_{2}={\rm Im}\left[ V^{\ast}_{11}V_{13}\right],
\end{equation} 
which in the standard parametrization of the PMNS matrix can be written in equivalent forms as follows:
\begin{equation}\label{IM2}
I_{1}=\cos{\theta_{12}}\sin{\theta_{12}}\cos^{2}{\theta_{13}}\sin{(\alpha/2)},\qquad I_{2}=\cos{\theta_{12}}\sin{\theta_{13}}\cos{\theta_{13}}\sin{(\beta/2-\delta_{CP}}). 
\end{equation}
Comparing these two equivalent definitions and taking into account  Eqs.~(\ref{pmnsen}), (\ref{mixan})  and  (\ref{fitpointneutrino})  we find for the Majorana phases 
\begin{eqnarray}\label{eq:Majorana-phases}
&&\alpha\simeq 32^{\circ}, \ \ \ \beta \simeq 55^{\circ}
\end{eqnarray}
for both the normal and inverted neutrino mass hierarchies.

\begin{figure}[h]
\centering
\includegraphics[width=0.5\textwidth]{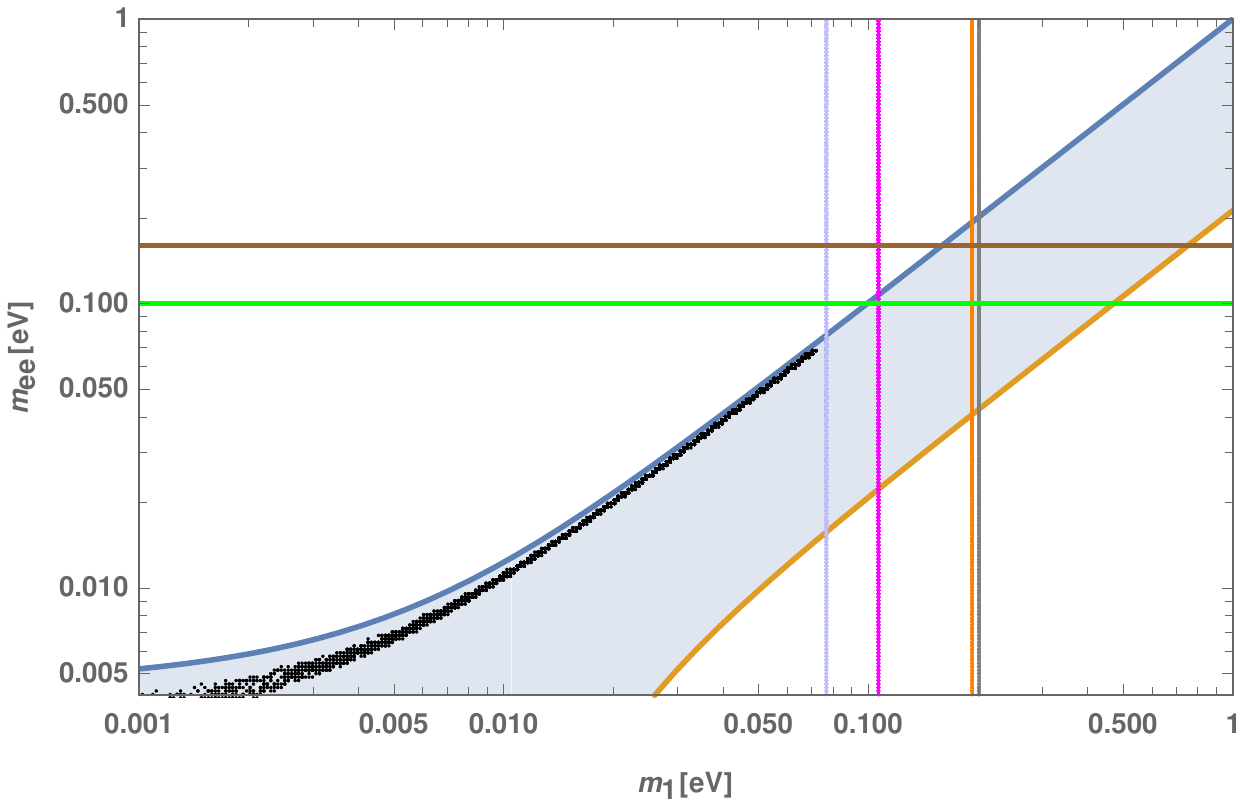}\includegraphics[width=0.5\textwidth]{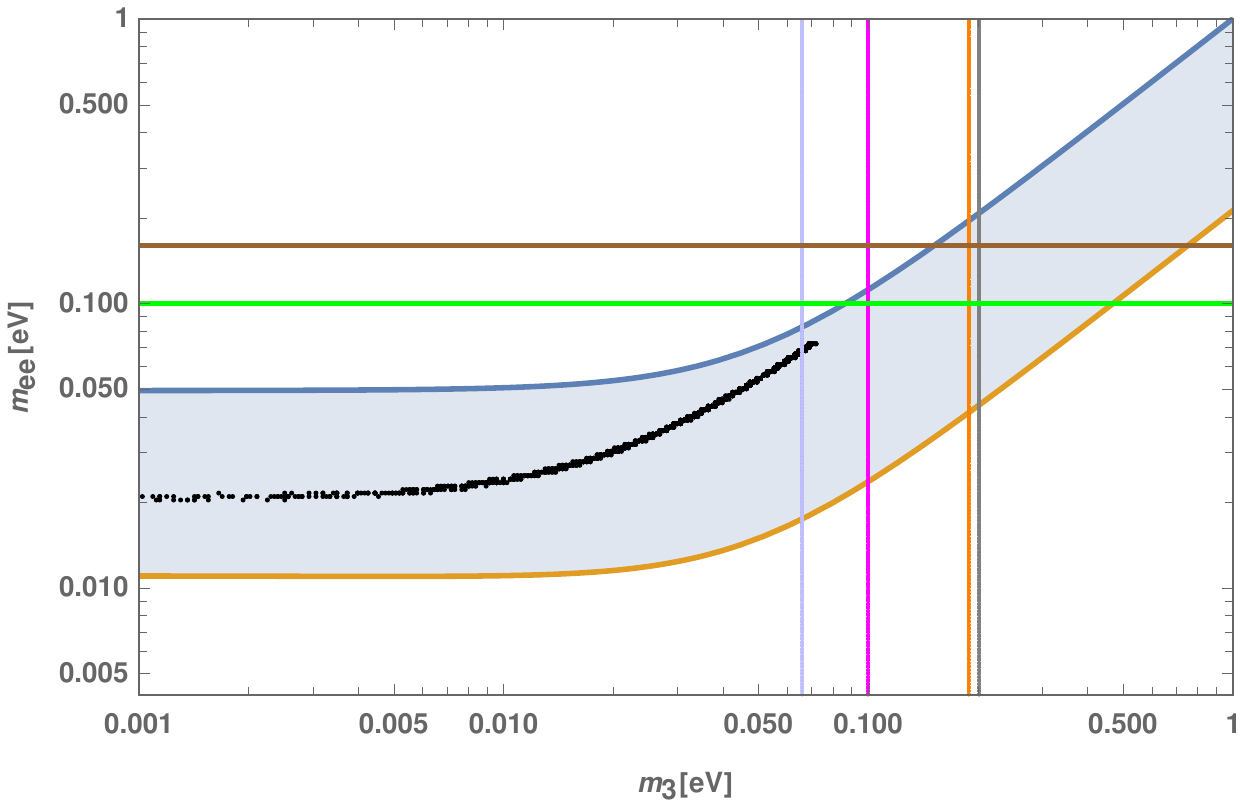}
\caption{Effective Majorana neutrino mass parameter $m_{ee}$ as a function of the lightest active neutrino mass $m_{1}$ for the normal (left pannel) and inverted (right pannel) neutrino mass hierarchies. The green and brown horizontal lines are the upper bounds $0.1$ eV and $0.16$ eV from the KamLAND-Zen \cite{KamLAND-Zen:2016pfg} and GERDA \textquotedblleft phase-II\textquotedblright \cite{Abt:2004yk,Ackermann:2012xja} experiments, respectively. The vertical blue, magenta, orange and gray lines are the upper bounds on the lightest neutrino mass from the combination of Cosmic Microwave  Background (CMB) radiation and Baryon acoustic oscillations (BAO) data \cite{Ade:2015xua},  Cosmic Microwave  Background (CMB) radiation as well as lensing observations \cite{Battye:2013xqa}, Wilkinson Microwave Anisotropy Probe (WMAP) Observations \cite{0067-0049-192-2-18} and Katrin experiment \cite{Wolf:2008hf}, respectively. The black curve corresponds to the predictions of our model.
}
\label{mee}
\end{figure}

Another important observable, to be considered, is the effective Majorana neutrino mass parameter of neutrinoless double beta decay ($0\nu\beta\beta$) defined as:
\begin{equation}\label{mef}
m_{ee} =\left| \sum_{i=1}^{3} m_{i} V_{1i}^{2}  \right|.
\end{equation} 
In our model the mass parameter, $m_{ee}$, depends on the two model parameters 
$\theta_{\nu}$, $\theta_{l}$ and  
lightest neutrino mass $m_{1}$ for the normal and $m_{3}$ for inverted hierarchy by virtue of 
\begin{eqnarray}\label{nmass12}
m_{2}&=&\sqrt{\Delta m^{2}_{21}+m^{2}_{1}},\hspace{16mm} m_{3}=\sqrt{\Delta m^{2}_{31}+m^{2}_{1}}\ \ \ \  \, \textrm{normal hierarchy},\nonumber\\
m_{2}&=&\sqrt{\Delta m^{2}_{13}+\Delta m^{2}_{21}+m^{2}_{3}},\hspace{3mm} 
m_{1}=\sqrt{\Delta m^{2}_{13}+m^{2}_{3}},\ \ \  \textrm{inverted hierarchy}.
\end{eqnarray}
In Figs.~\ref{mee} we show the conventional plots for correlations between the effective Majorana neutrino mass parameter $m_{ee}$ and the lightest active neutrino mass
 for the normal (left plot) and inverted (right plot) neutrino mass hierarchies. The arrays of black points in  these Figures were generated in our model by randomly variating the model parameters 
 $\theta_{\nu}$ and $\theta_{l}$ 
around their best-fit values (\ref{fitpointneutrino}) inside the $3\sigma$ experimental range of the mixing angles $\theta_{ij}$ (see Tables~\ref{neutrinoDataNH},\ref{neutrinoDataIH}). 
The lightest active neutrino mass was randomly varied in the range 
$0 < m_{1}< 0.2$ eV for the normal and $0 < m_{3}< 0.1$ eV for the inverted neutrino mass hierarchy. As seen from Figs.~\ref{mee} the effective Majorana neutrino mass parameter $m_{ee}$
lies within the limited ranges. Numerically they are
\begin{eqnarray}\label{eq:mbb-1}
 0.001\, \mbox{eV} &\leq& m_{ee} \leq  0.05\,\mbox{eV}\hspace{3mm} \mbox{NH}\\
 0.02\ \mbox{eV} &\leq& m_{ee} \leq 0.05\,\mbox{eV}\hspace{3mm} \mbox{IH}.
\end{eqnarray}
Note that in our model the parameter $m_{ee}$ is bounded from below even in the case of normal hierarchy.
These values are within the declared reach of the next-generation
bolometric CUORE experiment \cite{Alessandria:2011rc} or, more
realistically, of the next-to-next-generation ton-scale
$0\nu \beta \beta $-decay experiments (for a recent review see for
instance Ref. \cite{Agostini:2017jim}).  The currently most stringent
experimental upper bound $m_{ee}\leq 160$ meV is set by
$T_{1/2}^{0\nu\beta \beta }(^{136}\mathrm{Xe})\geq 1.1\times 10^{26}$
yr at 90\% C.L.  from the KamLAND-Zen experiment
\cite{KamLAND-Zen:2016pfg}.

Finally, let us briefly comment on the lepton flavor violating
process, $\mu \rightarrow e~\gamma$, which may be phenomenologically
dangerous for the models with the SM triplet scalars
\cite{Perez:2008zc, Fukuyama:2009xk, Akeroyd:2009nu,
  Melfo:2011nx}. Our model is this very case: it has the triplet
$\Delta$ with the decomposition shown in Eq.~(\ref{eq:VEVs-1}).
The contribution of the singly $\Delta^{+}$ and doubly $\Delta^{++}$ charged scalars to
the branching ratio (BR) of this process was found in Ref.~\cite{Lindner:2016bgg}.  In our model this contribution can be written in the form
%
\begin{equation}
\label{eq:Mu-EGamma-1}
{\rm BR} \left(\mu\rightarrow e~\gamma \right)\approx 4.5\times 10^{-3}\left(\frac{1}{\sqrt{2}v_{\Delta}\lambda^{9}}\right)^{4}\left| \left({\bf V}^{\ast}{\bf \hat{M}}^{\dagger}_{\nu}{\bf \hat{M}}_{\nu}{\bf V}^{T} \right)_{e\mu}\right|^{2}
\left(\frac{200~GeV}{m_{\Delta^{++}}}\right)^{4}
\end{equation}
where 
${\bf V}$ is the PMNS mixing matrix. We assumed for simplicity 
$m_{\Delta^{+}} = m_{\Delta^{++}}$ for the masses of the singly and doubly charged scalars. 
Note that the PMNS matrix elements, depending in our model only on the $\theta_{\nu}$ and $\theta_{l}$ parameters, have already been fixed numerically (\ref{fitpointneutrino})  from the fit to the experimental data on the neutrino mixing angles.
The neutrino masses are related by Eq. (\ref{nmass12}) to the lightest neutrino mass $m_{1}$ or $m_{3}$ for NH or IH, respectively.
From Eq.~(\ref{eq:Mu-EGamma-1}) we find the model prediction 
\begin{eqnarray}\label{eq:MU-E-Gamma-model-lim}
&&{\rm BR}\left(\mu \rightarrow e~\gamma\right)< 1.1\times 10^{-23}
\ \mbox{Model}
\end{eqnarray}
for the range of the parameters: $0 \lesssim m_{1}\lesssim 0.2$ eV for NH and 
$0 \lesssim m_{3}\lesssim 0.1$ eV for IH as well as
\mbox{$80$ Gev $< m^{++}_{\Delta}$} GeV 
and  \mbox{$v_{\Delta}<5$ GeV}. 
The latter upper bound comes from the precision measurements of the SM $\rho$-parameter. 
The lower limit for the mass of the singly and doubly charged Higges $m^{++}_{\Delta}$ derives from various low and high energy data \cite{Patrignani:2016xqp}.
The current experimental bound 
BR$\left(\mu \rightarrow e~\gamma\right)< 4.2\times 10^{-13}$ 
\cite{Patrignani:2016xqp} is significantly weaker than the model prediction 
(\ref{eq:MU-E-Gamma-model-lim}). 
Therefore, the model easily passes this phenomenological test.

\section{Conclusions}

\label{conclusions} 
We constructed  a multiscalar singlet extension of the singlet-triplet Higgs model, which explains the observed pattern of the quark and lepton masses and mixing by imposed symmetries and the field content.
The model incorporates the $\Delta \left( 27\right) $ family symmetry, which is
supplemented by the $Z_{16}\times Z_{24}$ discrete group. The observed
hierarchy of SM charged fermion masses and mixing angles is produced by the spontaneous breaking of the $\Delta \left( 27\right) \times Z_{16}\times Z_{24}$
discrete group at a very high energy-scale. The light active neutrino masses arise
from a type-II seesaw mechanism mediated by the neutral component of the 
$SU(2)_{L}$ scalar triplet. The obtained physical observables for both quark
and lepton sectors are compatible with their experimental values. In the quark
sector, there is a particular scenario inspired by naturalness arguments, which allowed us
to reduce the number of the model parameters from ten to four.  Even in this restricted scenario the model was able to accurately fit the ten observables of the SM quark sector. 
Moreover we found in the model a natural benchmark scenario with just two free parameters allowing to reasonably reproduce the experimental values of these ten observables.

The model is also successful in the neutrino sector,  reproducing the mixing angles within 
2 and 3$\sigma$ of
their most recent experimental values. The inverted hierarchy is slightly favored by the model.
The model predicts the effective Majorana neutrino mass parameter of neutrinoless
double beta decay in the range $0.001(0.02)$ eV $\lesssim m_{ee}\lesssim 0.05$ eV for the normal (inverted) neutrino spectrum. We found the Jarlskog invariant and leptonic Dirac CP violating phase in the ranges $-3.0\times 10^{-2}\lesssim J_{CP}\lesssim -2.40\times 10^{-2}$ and $230^{\circ}\lesssim\delta_{CP}\lesssim 236^{\circ}$  both for normal and inverted neutrino mass spectrum.
In principle, we can also accommodate dark matter, adding to our model two complex scalar singlets, $\Phi_1$ and $\Phi_2$ with lepton number $L=1$, as was done in Ref. \cite{Ma:2017xxj}. In this work, we do not pretend to address the dark matter problem, which is beyond the scope of this paper and will be addressed elsewhere.

\section*{Acknowledgments}

This research has received funding from Chilean grants Fondecyt No. 1170803,
No. 1150792 and CONICYT PIA/Basal FB0821 and ACT1406; and Mexican grant
237004, PAPIIT IN111518. JCGI thanks CINVESTAV for the warm hospitality and Red de Altas Energ\'{\i}as-CONACYT for the financial
support.

\appendix
\section{The product rules of the $\Delta (27)$ discrete group}

\label{A}

The $\Delta (27)$ discrete group is a subgroup of $SU(3)$ having 27 elements
divided into 11 conjugacy classes. It 
has the following 11 irreducible representations: one triplet $\mathbf{3}$, one antitriplet $\overline{\mathbf{3}}$ and nine singlets  $\mathbf{1}_{k,l}$ ($k,l=0,1,2$), where $k$ and $l$ correspond to the $Z_{3}$ and $Z_{3}^{\prime }$ charges, respectively.
\begin{eqnarray}
\mathbf{3}\otimes\mathbf{3}&=&\overline{\mathbf{3}}_{S_1}\oplus\overline{\mathbf{3}}_{S_2}\oplus\overline{\mathbf{3}}_{A}\notag\\
\overline{\mathbf{3}}\otimes\overline{\mathbf{3}}&=&\mathbf{3}_{S_1}\otimes\mathbf{3}_{S_2}\oplus\mathbf{3}_{A}\notag\\
\mathbf{3}\otimes\overline{\mathbf{3}}&=&\sum^{2}_{r=0}\mathbf{1}_{r,0}\oplus\sum^{2}_{r=0}\mathbf{1}_{r,1}\oplus\sum^{2}_{r=0}\mathbf{1}_{r,2}\notag\\
\mathbf{1}_{k,\ell }\otimes\mathbf{1}_{k^{\prime },\ell ^{\prime }}&=&\mathbf{1}_{k+k^{\prime }mod 3,\ell +\ell ^{\prime }mod 3}\\
\end{eqnarray}
Denoting $\left( x_{1},y_{1},z_{1}\right) $ and $\left(x_{2},y_{2},z_{2}\right) $ as the basis vectors for two  $\Delta(27)$-triplets $\mathbf{3}$, one finds:
\begin{eqnarray}\label{triplet-vectors}
\left( \mathbf{3}\otimes \mathbf{3}\right) _{\overline{\mathbf{3}}_{S_1}}&=&\left(x_{1}y_{1},x_{2}y_{2},x_{3}y_{3}\right),\notag\\
\left( \mathbf{3}\otimes \mathbf{3}\right)_{\overline{\mathbf{3}}_{S_2}}&=&\frac1{2}\left(x_{2}y_{3}+x_{3}y_{2},x_{3}y_{1}+x_{1}y_{3},x_{1}y_{2}+x_{2}y_{1}\right) ,\notag\\
\left( \mathbf{3}\otimes \mathbf{3}\right)_{\overline{\mathbf{3}}_{A}}&=&\frac1{2}\left(x_{2}y_{3}-x_{3}y_{2},x_{3}y_{1}-x_{1}y_{3},x_{1}y_{2}-x_{2}y_{1}\right)
,\notag\\ 
\left( \mathbf{3}\otimes\overline{\mathbf{3}}\right)_{\mathbf{1}_{r,0}}&=&x_{1}y_{1}+\omega^{2r}x_{2}y_{2}+\omega^{r}x_{3}y_{3},\notag\\
\left( \mathbf{3}\otimes\overline{\mathbf{3}}\right)_{\mathbf{1}_{r,1}}&=&x_{1}y_{2}+\omega^{2r} x_{2}y_{3}+\omega^{r}x_{3}y_{1},\notag\\
\left( \mathbf{3}\otimes\overline{\mathbf{3}}\right)_{\mathbf{1}_{r,2}}&=&x_{1}y_{3}+\omega^{2r}x_{2}y_{1}+\omega^{r}x_{3}y_{2},
\end{eqnarray}
where $r=0,1,2$ and $\omega =e^{i \frac{ 2 \pi }{3}}$. More details on the $\Delta (27)$ discrete group are provided in Refs. \cite{Ishimori:2010au,Hernandez:2016eod,Vien:2016tmh}

\end{document}